\documentclass[11pt,onecolumn,draftcls]{IEEEtran} \newcommand{\figsize}{3.5in}
\usepackage{graphicx,psfrag}
\usepackage[usenames]{color}
\usepackage[cmex10]{amsmath}
\usepackage{amsfonts,amssymb,latexsym,ifthen,color}
\newboolean{SEP_FIG_CAPS}\setboolean{SEP_FIG_CAPS}{false}

\ifthenelse{\boolean{SEP_FIG_CAPS}}
{
 \newcommand{\putFrag}[4]{\begin{figure}[p]
                            \centering
                            #4
			    \includegraphics[width=#3]{figures/#1.eps}
            		    \caption{}
     			    \label{fig:#1}
                          \end{figure}
                          \clearpage}
 \newcommand{\putTable}[3]{\begin{table}[p]
  			    \centering
		            #3
            		    \caption{}
     			    \label{tab:#1}
			  \end{table}
			  \clearpage}
 \newcommand{\capFrag}[2]{\noindent Fig.~\ref{fig:#1}. #2 \medskip\\}
 \newcommand{\capTable}[2]{\noindent Tab.~\ref{tab:#1}. #2 \medskip\\}
}
{
 \newcommand{\putFrag}[4]{\begin{figure}[t]
                            \centering
                            #4
			    \includegraphics[width=#3]{#1.eps}
            		    \caption{#2}
           		    \label{fig:#1}
                          \end{figure} }
 \newcommand{\putTable}[3]{\begin{table}[t]
  			    \centering
		            #3
     			    \caption{#2}
     			    \label{tab:#1}
			  \end{table} }
 \newcommand{\capFrag}[2]{}
 \newcommand{\capTable}[2]{}
}

 \newcommand{\defn}{\triangleq}

 \newcommand{\hvec}[1]{\ensuremath{\Hat{\boldsymbol{#1}}}}
    
 \renewcommand{\vec}[1]{\ensuremath{\boldsymbol{#1}}}

 \newcommand{\norm}[1]{\ensuremath{\| #1 \|}}
 \newcommand{\mc}[1]{\ensuremath{\mathcal{#1}}}

 \newcommand{\Complex}{{\mathbb{C}}}
 \newcommand{\Int}{{\mathbb{Z}}}

 \newcommand{\tran}{^\textsf{T}}
 
 \newcommand{\of}[1]{^{(#1)}}

 \DeclareMathOperator{\E}{E}
 \DeclareMathOperator{\var}{var}

 \DeclareMathOperator{\polylog}{polylog}

 \renewcommand{\eqref}[1]{(\ref{eq:#1})}
 
 \newcommand{\Figref}[1]{Figure~\ref{fig:#1}}
 \newcommand{\figref}[1]{Fig.~\ref{fig:#1}}
 \newcommand{\tabref}[1]{Table~\ref{tab:#1}}
 \newcommand{\secref}[1]{Section~\ref{sec:#1}}
 
 \newcommand{\appref}[1]{Appendix~\ref{app:#1}}

 \newcommand{\textr}[1]{\textcolor{Red}{#1}}
 \newcommand{\textg}[1]{\textcolor{Green}{#1}}
 
 \newcommand{\texto}[1]{\textcolor{Orange}{#1}}
 \newcommand{\textc}[1]{\textcolor{Cyan}{#1}}

 \newcounter{comment}[section]
 
 \newcounter{texthead}[section]

\newcommand{\giv}{\,|\,}

\newcommand{\const}{\mathbb{S}}

\newcommand{\Lhpd}{L_{\textsf{hpd}}}
\newcommand{\Np}{N_\textsf{p}}
\newcommand{\Nd}{N_\textsf{d}}
\newcommand{\Md}{M_\textsf{d}}
\newcommand{\Mc}{M_\textsf{c}}
\newcommand{\Mi}{M_\textsf{i}}
\newcommand{\Mt}{M_\textsf{t}}
\newcommand{\pt}{_\textsf{pt}}
\newcommand{\SNR}{\textsf{SNR}}
\newcommand{\BER}{\textsf{BER}}
\newcommand{\NMSE}{\textsf{NMSE}}
\newcommand{\Ls}{K}

\newcommand{\inp}{_{\textsf{in},j}}
\newcommand{\out}{_{\textsf{out},i}}
\newcommand{\sparse}{_\text{\sf sparse}}
\newcommand{\nonsparse}{_\text{\sf non-sparse}}

\begin{document}
\setlength{\arraycolsep}{0.8mm}
 \title{Belief-propagation-based joint channel estimation and decoding 
	for spectrally efficient communication over unknown sparse channels}
 \author{Philip Schniter\IEEEauthorrefmark{1}%
        \thanks{The author is with the 
                Department of Electrical and Computer Engineering
                at The Ohio State University, Columbus, OH.}%
        \thanks{Please direct all correspondence to 
                Prof.\ Philip Schniter,
                Dept. ECE, 2015 Neil Ave., Columbus OH 43210,
                e-mail: schniter@ece.osu.edu,
                phone 614.247.6488, fax 614.292.7596.}%
        \thanks{This work was supported in part by 
                the National Science Foundation grant CCF-1018368
                and DARPA/ONR grant N66001-10-1-4090.}
        \thanks{Portions of this work were presented at the 
		Asilomar 2010 Conference on Signals, Systems, and Computers.}%
        }
 \date{\today}
\maketitle

\begin{abstract}
We consider spectrally-efficient communication over a Rayleigh 
$N$-block-fading channel with a $\Ls$-sparse $L$-length discrete-time 
impulse response (for $0\!<\!\Ls\!<\!L\!<\!N$), where neither the 
transmitter nor receiver know the channel's coefficients nor its support.
Since the high-$\SNR$ ergodic capacity of this channel has been shown 
to obey $C(\SNR)=(1\!-\!\Ls/N)\log_2(\SNR)+\mc{O}(1)$, any pilot-aided
scheme that sacrifices more than $\Ls$ dimensions per fading block
to pilots will be spectrally inefficient.
This causes concern about the conventional ``compressed channel sensing''
approach, which uses $\mc{O}\big(\Ls \polylog(L) \big)$ pilots.
In this paper, we demonstrate that practical spectrally-efficient 
communication is indeed possible.
For this, we propose a novel belief-propagation-based reception 
scheme to use with a standard bit-interleaved coded 
orthogonal frequency division multiplexing (OFDM) transmitter.
In particular, we leverage the ``relaxed belief propagation''
methodology, which allows us to perform joint sparse-channel
estimation and data decoding with only $\mc{O}(LN)$ complexity.
Empirical results show that our receiver achieves the desired capacity 
pre-log factor of $1-\Ls/N$ and performs near genie-aided bounds 
at both low and high $\SNR$.

\end{abstract}

\section{Introduction} 				\label{sec:intro}

Our goal is to communicate, in a spectrally efficient manner, over 
a Rayleigh $N$-block-fading channel with a $\Ls$-sparse discrete-time 
impulse response of length $L$ (where $0<\Ls<L<N$), 
under the realistic assumption that neither the transmitter nor the receiver
knows the channel's coefficients nor its support.
It has been recently shown \cite{Kannu:arXiv:10} that the ergodic capacity 
of this noncoherent sparse channel obeys 
\begin{eqnarray}
  C\sparse(\SNR) 
  &=& \frac{N-\Ls}{N}\log_2(\SNR) + \mc{O}(1) 	\label{eq:cap}
\end{eqnarray}
as the signal-to-noise ratio ($\SNR$) grows large. 
For comparison, the high-$\SNR$ ergodic noncoherent capacity of the 
Rayleigh $N$-block-fading $L$-length \emph{non}-sparse channel obeys 
\cite{Vikalo:TSP:04}
\begin{eqnarray}
  C\nonsparse(\SNR) 
  &=& \frac{N-L}{N}\log_2(\SNR) + \mc{O}(1) ,	\label{eq:cap2}
\end{eqnarray}
which exhibits a lower pre-log factor than \eqref{cap}.
Thus, information theory confirms that channel sparsity can indeed be 
exploited to increase spectral efficiency, at least for high $\SNR$.
In particular, it establishes that, in the high-$\SNR$ regime, 
the signaling scheme does not need to sacrifice more than $\Ls$ 
degrees-of-freedom per fading-block to mitigate the effects of not knowing
the $\Ls$ non-zero channel coefficients nor their locations.
%

Among the many strategies that exist for communication over unknown
channels, pilot-aided transmission (PAT) \cite{Tong:SPM:04}
has emerged as one of the most effective.
For example, it is known \cite{Vikalo:TSP:04} that, for
the Rayleigh $N$-block-fading $L$-length \emph{non}-sparse channel,
PAT achieves rates in accordance with the capacity expression
\eqref{cap2}.\footnote{
  Note that \eqref{cap} and \eqref{cap2} specify only that the
  maximum rate of reliable communication grows in linear proportion 
  to $\log(\SNR)$ according to the specified pre-log factor;
  the exact value of the capacity remains unspecified due to 
  the $\mc{O}(1)$ term.}
It is then not surprising that the vast majority of techniques that have 
recently been proposed for communication over \emph{sparse} channels are 
also based on PAT 
(see, e.g., the extensive bibliography in \cite{Bajwa:PROC:10}).
Broadly speaking, these techniques propose to exploit channel sparsity in 
order to reduce
the number of pilots used for accurate channel estimation, with the 
end goal of increasing spectral efficiency.
Typically, these schemes take a \emph{decoupled} approach to reception:
a sparse-channel estimate is calculated from pilot observations 
using a practical compressed sensing algorithm like LASSO
\cite{Tibshirani:JRSSb:96,Chen:JSC:98},
and the channel estimate is subsequently used for data decoding.
Hereafter, we shall refer to this decoupled approach 
as ``\emph{compressed channel sensing}'' (CCS), after \cite{Bajwa:PROC:10}. 
When $\mc{O}\big(\Ls \polylog(L) \big)$ pilots\footnote{
	The use of $\mc{O}\big(\Ls \polylog(L) \big)$ pilots corresponds to
	the case of OFDM-based transmission, which is the case
	that we focus on later in this paper.} 
are used for CCS,
the theory of \emph{compressed sensing} guarantees that---with high
probability---the resulting channel estimates will be accurate, 
e.g., their squared-error will decrease in proportion to the 
received noise variance \cite{Bajwa:PROC:10}.

While the use of $\mc{O}\big(\Ls \polylog(L) \big)$ pilots may be 
an improvement over $L$ pilots required when channel sparsity is not
taken into account, the capacity expression \eqref{cap} implies that
any PAT scheme sacrificing more than $\Ls$ degrees of freedom (per fading block)
to pilots will be spectrally inefficient in the high-$\SNR$ regime.
Thus, any scheme based on CCS, which uses 
$\mc{O}\big(\Ls \polylog(L) \big) > \Ls$ pilots, will fall short of
maximizing spectral efficiency.
One may then wonder whether there exists a practical\footnote{
  In \cite{Kannu:arXiv:10}, a scheme that achieves the prelog factor in 
  \eqref{cap} was proposed, but it is impractical in the sense that 
  its complexity grows exponentially with the fading-block length $N$.}
communication scheme that achieves the capacity prelog factor in \eqref{cap}.

In this paper, we propose a novel approach to communication over
sparse channels that (empirically) achieves rates 
in accordance with the sparse-channel capacity expression \eqref{cap}.
For transmission, we use a conventional scheme, based on 
bit-interleaved coded modulation (BICM) 
with orthogonal frequency division multiplexing (OFDM) 
and a few carefully placed training bits. 
For reception, we deviate from the CCS approach and perform 
sparse-channel estimation and data decoding \emph{jointly}.
To accomplish this latter task in a practical manner, we take an 
approach suggested by \emph{belief-propagation} (BP) \cite{Pearl:Book:88},
leveraging recent advances in ``relaxed BP'' \cite{Guo:ISIT:07,Rangan:10v2} 
and in BP-based soft-input/soft-output (SISO) decoding \cite{MacKay:Book:03}.
The scheme that we propose has very low computation complexity: 
only $\mc{O}(NL)$ multiplies per fading block are required.
Thus, we are able to handle long channels, many subcarriers, and 
large QAM constellations (which are in fact necessary to achieve 
high spectral efficiencies). 
Our simulations, for example, use $N\!=\!1021$ subcarriers, up to $256$-point 
QAM constellations, $\approx\! 10000$-bit LDPC codes, and channels with 
length $L\!=\!256$ and average sparsity $\E\{\Ls\}\!=\!64$. 
Under these conditions, we find that our scheme yields error rates that
are close to genie-aided bounds, and far superior to CCS, 
in both low- and high-$\SNR$ regimes.
Moreover, we find that the outage rate behavior of our scheme 
coincides with the sparse-channel capacity expression \eqref{cap}.

We will now place our work in context.
The basic idea of using BP for joint channel-estimation 
and decoding (JCED) has been around for more than a decade 
(see, e.g., the early overview \cite{Worthen:TIT:01} and the
more recent works \cite{Jin:TCOM:08,Guo:JSAC:08}). 
The standard rules of BP specify that 
messages are passed among nodes of the factor graph according to the 
sum-product algorithm (SPA) \cite{Pearl:Book:88}.
However, since in many cases exact implementation of SPA on the JCED 
factor graph is impractical, SPA must be approximated, and there is 
considerable freedom as to how this can be done.
In fact, many well known iterative estimation algorithms can be 
recognized as particular approximations of SPA-BP:
the expectation-maximization (EM) algorithm 
\cite{Dauwels:ISIT:05},
particle filtering \cite{Dauwels:ISIT:06},
variational Bayes (or ``mean-field'') \cite{Dauwels:ISIT:07},
and even steepest descent \cite{Dauwels:ITW:05}.
Not surprisingly, this plurality of possibilities has yielded 
numerous BP-based JCED designs for frequency-selective channels 
(e.g., \cite{Novak:ICASSP:09,Liu:PIMRC:09,Knievel:ICC:10,Kirkelund:GLOBE:10}).
%
%

Our work is distinct from the existing BP-based JCED literature in that
1) we model the channel as apriori sparse (i.e., the coefficients are 
non-Gaussian) whereas, in all of the existing BP-based JCED work that 
we are aware, the channel coefficients\footnote{
  After submitting this manuscript, we became aware of the related 
  work \cite{Zhu:TWC:09}, which applies BP to JCED for flat-fading Gaussian 
  channel coefficients and non-Gaussian interference.} 
are modeled as Gaussian, and 
2) we leverage a state-of-the-art BP approximation known as 
``relaxed BP'' (RBP), which has been rigorously analyzed and shown to 
yield asymptotically exact posteriors (as the problem dimensions
$N,L\rightarrow\infty$ and under certain technical assumptions on
the mixing matrix) \cite{Guo:ISIT:07,Rangan:10v2}.
In fact, we conjecture that the success of our method is due in large 
part to the principled approximations used within RBP.
We also note that, although we focus on the case of sparse channels,
our approach would be applicable to non-sparse channels or, e.g.,
non-sparse channels with unknown length \cite{Novak:ICASSP:09},
with minor modification of the assumed channel priors.

Our paper is organized as follows.
In \secref{model} we detail the system model, and
in \secref{albp} we detail our RBP-based JCED approach.
In \secref{sims} we report the results of our simulation study, and 
in \secref{conc} we conclude.

\section{System Model} 				\label{sec:model}

We assume an OFDM-based transmitter that uses a total of $N$ subcarriers, 
each modulated by a QAM symbol from a $2^M$-ary unit-energy 
constellation $\const$.
Of these subcarriers, $\Np$ are dedicated as pilots,\footnote{
  For our BP-based JCED, we will see in \secref{sims} that 
  $(\Np,\Mt)\!=\!(0,M\Ls)$ is most effective.}
and the remaining
$\Nd\!\defn\! N-\Np$ are used to transmit a total of $M_t$ training bits 
and $\Md\! \defn\! \Nd M \!-\! \Mt$ coded/interleaved data bits.
To generate the latter, we encode $\Mi$ information bits using a
rate-$R$ coder, interleave them, and partition the resulting 
$\Mc\!\defn\!\Mi/R$ 
bits among an integer number $T\!\defn\!\Mc/\Md$ of OFDM symbols.
The resulting scheme has a spectral efficiency of 
$\eta\! \defn\! \Md R/N$ information bits per channel use (bpcu).
It should be emphasized that our model supports both subcarriers whose QAM 
symbols are completely known to the receiver (``pilot subcarriers''), 
as well as subcarriers whose QAM symbols are only partially known to the 
receiver (via ``training bits''). 
In our nomenclature, the known bits that make up a ``pilot subcarrier'' 
are distinct from the ``training bits'' that may be sprinkled among
the ``data subcarriers.''

In the sequel, 
we use $s\of{k}\in\const$ for $k\in\{1,\dots,2^M\}$ to denote the 
$k^{th}$ element of the QAM constellation, and 
$\vec{c}\of{k}\!\defn\! (c_{1}\of{k},\dots,c_{M}\of{k})\tran\in\{0,1\}^M$ 
to denote the 
bits corresponding to $s\of{k}$ as defined by the symbol mapping.
Likewise, we use $s_i[t]\in\const$ to denote the QAM symbol transmitted 
on the $i^{th}$ subcarrier of the $t^{th}$ OFDM symbol and 
$\vec{c}_i[t] \!\defn\! (c_{i,1}[t],\dots,c_{i,M}[t])\tran\in\{0,1\}^M$ 
to denote the 
(coded/interleaved or training or pilot) bits corresponding to that symbol.
We then collect the $NM$ bits that make up the $t^{th}$ OFDM symbol into 
$\vec{c}[t] \!\defn\! (\vec{c}_0[t],\dots,\vec{c}_{N-1}[t])\tran$, and
we collect the $NMT$ bits that make up the entire (interleaved) codeword into
$\vec{c} \!\defn\! (\vec{c}[1],\dots,\vec{c}[T])\tran\in\{0,1\}^{TNM}$.
The elements of $\vec{c}$ that are known apriori as pilot or training bits 
will be referred to as $\vec{c}\pt$.
The remainder of $\vec{c}$ is determined from the information
bits $\vec{b}\!=\!(b_1,\dots,b_{\Mi})\tran$ by coding/interleaving.

We use the standard OFDM model (see, e.g., \cite{Cimini:TCOM:85})
for the received value on subcarrier $i$ of OFDM-symbol $t$: 
\begin{eqnarray}
  y_i[t]
  &=& s_i[t] z_i[t] + v_i[t] ,				\label{eq:yi}
\end{eqnarray}
where $z_i[t]\in\Complex$ denotes the $i^{th}$ subcarrier's gain and 
$\{v_i[t]\}$ denotes circular white Gaussian noise with variance $\mu^v$.
As usual, the subcarrier gains 
$\vec{z}[t]\defn (z_0[t],\dots,z_{N-1}[t])\tran$ 
are related to the baud-spaced channel impulse response vector
$\vec{x}[t]\defn (x_0[t],\dots,x_{L-1}[t])\tran$ via
$z_i[t] = \sum_{j=0}^{L-1} \Phi_{ij} x_j[t]$,
where $\Phi_{ij}=e^{-\sqrt{-1}\frac{2\pi}{N}ij}$ can be recognized as
the $(i,j)^{th}$ element of the $N$-DFT matrix $\vec{\Phi}$.
Throughout, we will use $j$ to index the lag of the impulse response.
We assume that the channel is block-fading with fading interval $N$, 
so that the vectors $\{\vec{x}[t]\}_{t=1}^T$ are i.i.d. across $t$.
To simplify our development of the algorithm, we assume in the sequel that 
$T=1$ and drop the ``$[t]$'' notation for brevity. 
However, for the simulations in \secref{sims}, we revert back to general 
$T$ in order to facilitate the use of long LDPC codewords.

As described in \secref{intro}, the focus of the paper is on 
block-Rayleigh-fading channels with sparse impulse responses 
$\{x_j\}$.
To model sparsity, we treat the impulse response coefficients as
random variables $\{X_j\}$ with the independent Bernoulli-Gaussian prior pdf.
\begin{eqnarray}
  p_{X_j}(x) 
  &=& \lambda_j \mc{CN}(x;0,\mu_j) 
	+ (1-\lambda_j)\delta(x),		\label{eq:pxj}
\end{eqnarray}
where $\mc{CN}(x;a,b) \!\defn\! (\pi b)^{-1} \exp(-b^{-1}|x-a|^2)$
denotes the complex-Gaussian pdf, 
$\delta(\cdot)$ the Dirac delta, 
$\lambda_j\!=\!\Pr\{X_j\!\neq\! 0\}$ the sparsity rate,
and $\mu_j\!=\!\var\{X_j\}$ the variance.
We furthermore assume that the channel is energy-preserving with  
an exponential delay-power profile, so that 
$\mu_j \!=\! 2^{-j/\Lhpd}/(\sum_{r=0}^{L-1}\lambda_r 2^{-r/\Lhpd})$,
where $\Lhpd$ denotes the half-power delay.
For simplicity, we assume a uniform sparsity rate of 
$\lambda=\lambda_j\,\forall j$.

The presence of a Dirac delta in \eqref{pxj} indicates that we assume an
``perfectly sparse'' channel model.
Although perfect sparsity is not expected to manifest in practice, it is
frequently assumed in the literature (see, e.g., \cite{Bajwa:PROC:10} and
the papers cited therein).
While the JCED algorithm proposed in \secref{jced} can handle generic
marginal priors $p_{X_j}(x)$,
we make the perfect sparsity assumption only to facilitate a direct comparison 
to the information theoretic result \eqref{cap} from \cite{Kannu:arXiv:10}.
In follow-on work \cite{Schniter:SPAWC:11,Schniter:JSTSP:11}, we
consider channel taps that are both clustered and non-perfectly sparse,
as motivated by the IEEE 802.15.4a model in combination with raised-cosine 
pulse shapes.

\section{BP-based Joint Channel Estimation and Decoding}	\label{sec:albp}

\putFrag{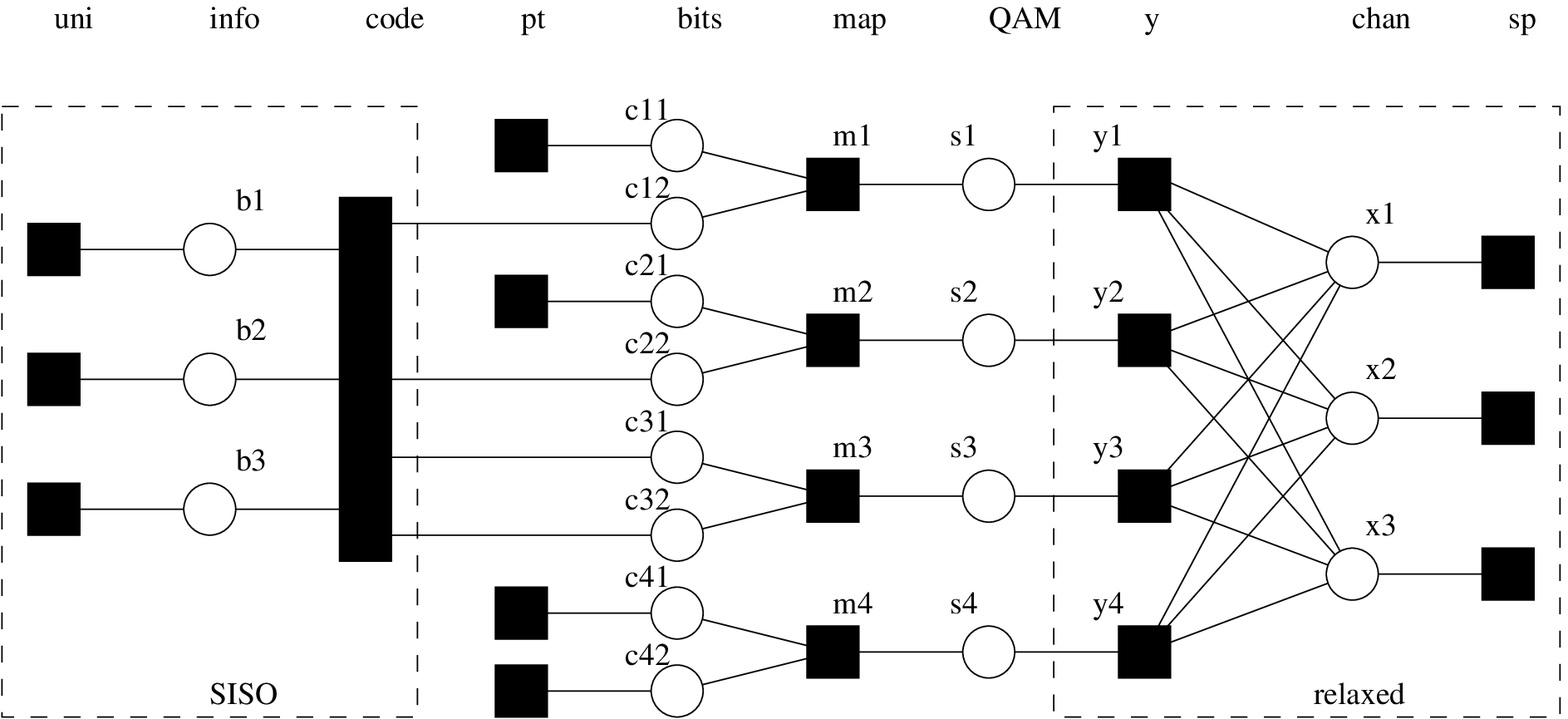}
	{
	 The factor graph of the JCED problem for a toy example with 
	 $N=4$ OFDM subcarriers, 
	 $M=2$ bits per QAM symbol, 
	 $\Mi=3$ information bits, 
	 $\Mt=2$ training bits, 
	 $\Np=1$ pilot subcarriers (at subcarrier index $i=3$), 
	 and a channel with length $L=3$.}
 	{3.4in}
	{\newcommand{\bs}{2.75}
  	 \newcommand{\vs}{0.8}
	 \newcommand{\cs}{0.7}
	 \newcommand{\ts}{0.55}
	 \psfrag{SISO}[B][Bl][\ts]{\sf SISO decoding}
	 \psfrag{relaxed}[B][Bl][\ts]{\sf RBP}
	 \psfrag{b1}[Bl][Bl][\vs]{$b_1$}
	 \psfrag{b2}[Bl][Bl][\vs]{$b_2$}
	 \psfrag{b3}[Bl][Bl][\vs]{$b_3$}
	 \psfrag{b4}[Bl][Bl][\vs]{$b_4$}
	 \psfrag{c11}[B][Bl][\cs]{$c_{0,1}$}
	 \psfrag{c21}[B][Bl][\cs]{$c_{1,1}$}
	 \psfrag{c31}[B][Bl][\cs]{$c_{2,1}$}
	 \psfrag{c41}[B][Bl][\cs]{$c_{3,1}$}
	 \psfrag{c12}[B][Bl][\cs]{$c_{0,2}$}
	 \psfrag{c22}[B][Bl][\cs]{$c_{1,2}$}
	 \psfrag{c32}[B][Bl][\cs]{$c_{2,2}$}
	 \psfrag{c42}[B][Bl][\cs]{$c_{3,2}$}
	 \psfrag{m1}[b][Bl][\vs]{$\mc{M}_0$}
	 \psfrag{m2}[b][Bl][\vs]{$\mc{M}_1$}
	 \psfrag{m3}[b][Bl][\vs]{$\mc{M}_2$}
	 \psfrag{m4}[b][Bl][\vs]{$\mc{M}_3$}
	 \psfrag{s1}[B][Bl][\vs]{$s_0$}
	 \psfrag{s2}[B][Bl][\vs]{$s_1$}
	 \psfrag{s3}[B][Bl][\vs]{$s_2$}
	 \psfrag{s4}[B][Bl][\vs]{$s_3$}
	 \psfrag{y1}[B][Bl][\vs]{$y_0$}
	 \psfrag{y2}[B][Bl][\vs]{$y_1$}
	 \psfrag{y3}[B][Bl][\vs]{$y_2$}
	 \psfrag{y4}[B][Bl][\vs]{$y_3$}
	 \psfrag{x1}[Bl][Bl][\vs]{$x_1$}
	 \psfrag{x2}[Bl][Bl][\vs]{$x_2$}
	 \psfrag{x3}[Bl][Bl][\vs]{$x_3$}
	 \psfrag{uni}[][Bl][\ts]{\sf
	 	\begin{tabular}{@{}c@{}}uniform\\[-\bs mm]prior\end{tabular}}
	 \psfrag{info}[][Bl][\ts]{\sf
	 	\begin{tabular}{@{}c@{}}info\\[-\bs mm]bits\end{tabular}}
	 \psfrag{code}[][Bl][\ts]{\sf
	 	\begin{tabular}{@{}c@{}}code \&\\[-\bs mm]interlv\end{tabular}}
	 \psfrag{pt}[][Bl][\ts]{\sf
	 	\begin{tabular}{@{}c@{}}pilots \&\\[-\bs mm]training\end{tabular}}
	 \psfrag{bits}[][Bl][\ts]{\sf
	 	\begin{tabular}{@{}c@{}}coded\\[-\bs mm]bits\end{tabular}}
	 \psfrag{map}[][Bl][\ts]{\sf
	 	\begin{tabular}{@{}c@{}}symbol\\[-\bs mm]mapping\end{tabular}}
	 \psfrag{QAM}[][Bl][\ts]{\sf
	 	\begin{tabular}{@{}c@{}}QAM\\[-\bs mm]symbs\end{tabular}}
	 \psfrag{y}[][Bl][\ts]{\sf
	 	\begin{tabular}{@{}c@{}}OFDM\\[-\bs mm]observ\end{tabular}}
	 \psfrag{chan}[][Bl][\ts]{\sf
	 	\begin{tabular}{@{}c@{}}impulse\\[-\bs mm]response\end{tabular}}
	 \psfrag{sp}[][Bl][\ts]{\sf
	 	\begin{tabular}{@{}c@{}}sparse\\[-\bs mm]prior\end{tabular}}
	 }

Our goal is to infer the information bits $\vec{b}$,
given the OFDM observations $\vec{y}\!\defn\!(y_0,\dots,y_{N-1})\tran$ 
and the pilot/training bits $\vec{c}\pt$,
in the absence of channel state information.
For simplicity, we assume that the channel statistics (i.e., 
$\{\mu^v,\lambda,\Lhpd,L\}$) are known.\footnote{
  Although it remains outside the scope of this work, it should be 
  possible to jointly estimate these statistics together with the 
  channel and data realizations by treating them as random variables
  with appropriate non-informative priors and expanding the factor 
  graph accordingly.}
In particular, we aim to maximize the posterior pmf 
$p(b_m\giv\vec{y},\vec{c}\pt)$ of each information bit $b_m$.
Given the model of \secref{model}, this posterior can be decomposed into
a product of factors as follows:
\begin{eqnarray}
  p(b_m\giv \vec{y},\vec{c}\pt) 
  &=& \sum_{\vec{b}_{\setminus m}} p(\vec{b}\giv \vec{y},\vec{c}\pt) 
  \,\propto\, 
  	\sum_{\vec{b}_{\setminus m}} p(\vec{y}\giv \vec{b},\vec{c}\pt) p(\vec{b}) 
  							\label{eq:propto} \\
  &=& \int_{\vec{x}} 
	\sum_{\vec{c}}
	\sum_{\vec{s}} 
  	\sum_{\vec{b}_{\setminus m}} 
  	p(\vec{y}\giv \vec{s}, \vec{x}) 
	p(\vec{x})
	p(\vec{s}\giv \vec{c})
	p(\vec{c}\giv \vec{b},\vec{c}\pt) p(\vec{b}) 	\\
  &=& \int_{\vec{x}} 
	\prod_{j=0}^{L-1} p(x_j)
	\sum_{\vec{s}} 
  	\prod_{i=0}^{N-1} p(y_i|s_i, \vec{x}) 
  	\sum_{\vec{c}}
	p(s_{i}|\vec{c}_{i})
  	\sum_{\vec{b}_{\setminus m}} 
	p(\vec{c}|\vec{b},\vec{c}\pt) 
	\prod_{m=1}^{\Mi} p(b_{m}),\quad	\label{eq:factored}
\end{eqnarray}
where ``$\propto$'' denotes equality up to a scaling and where
$\vec{b}_{\setminus m}$
denotes the vector $\vec{b}$ with the $m^{th}$ element omitted.
The factorization \eqref{factored} is illustrated by the \emph{factor 
graph} in \figref{factor_graph_noncoh}, where the round nodes represent
random variables and the square nodes represent the factors of the 
posterior identified in \eqref{factored}.

\subsection{Background on belief propagation}		\label{sec:bp}

While exact evaluation of the posteriors 
$\{p(b_m\giv\vec{y},\vec{c}\pt)\}_{m=1}^{\Mi}$
is computationally impractical for the problem sizes of interest,
these posteriors can be approximately evaluated using \emph{belief propagation}
(BP) \cite{Pearl:Book:88} on a loopy factor graph like the illustrated in 
\figref{factor_graph_noncoh}.
In standard BP, beliefs take the form of pdfs/pmfs that are propagated among 
nodes of the factor graph according to the rules of the 
\emph{sum-product algorithm} (SPA):
\begin{enumerate}
 \item
 If factor node $f(v_1,\dots,v_A)$ is connected to variable nodes 
 $\{v_a\}_{a=1}^A$, then the belief passed from $f$ to $v_b$ is 
 \begin{eqnarray}
   p_{f\rightarrow v_b}(v_b) 
   &\propto& \int_{\{v_a\}_{a\neq b}} f(v_1,\dots,v_A) 
 \prod_{a\neq b} p_{v_a \rightarrow f}(v_a),
 \end{eqnarray}
 where $\{p_{v_a\rightarrow f}(\cdot)\}_{a\neq b}$ are the
 beliefs most recently passed to $f$ from $\{v_a\}_{a\neq b}$.
 \item
 If variable node $v$ is connected to factor nodes $\{f_1,\dots,f_B\}$,
 then the belief passed from $v$ to $f_a$ is 
 \begin{eqnarray}
   p_{v\rightarrow f_a}(v)
   &\propto& \prod_{b\neq a} p_{f_b\rightarrow v}(v),
 \end{eqnarray}
 where $\{p_{f_b\rightarrow v}(\cdot)\}_{b\neq a}$ are the 
 beliefs most recently passed to $v$ from $\{f_b\}_{b\neq a}$.
 \item
 If variable node $v$ is connected to factor nodes $\{f_1,\dots,f_B\}$,
 then the posterior on $v$ is the product of all most recently 
 arriving beliefs, i.e., 
 \begin{eqnarray}
   p(v) 
   &\propto& \prod_{b=1}^B p_{f_b\rightarrow v}(v).
 \end{eqnarray}
\end{enumerate}
\Figref{bp} is provided to illustrate the first two rules.

\putFrag{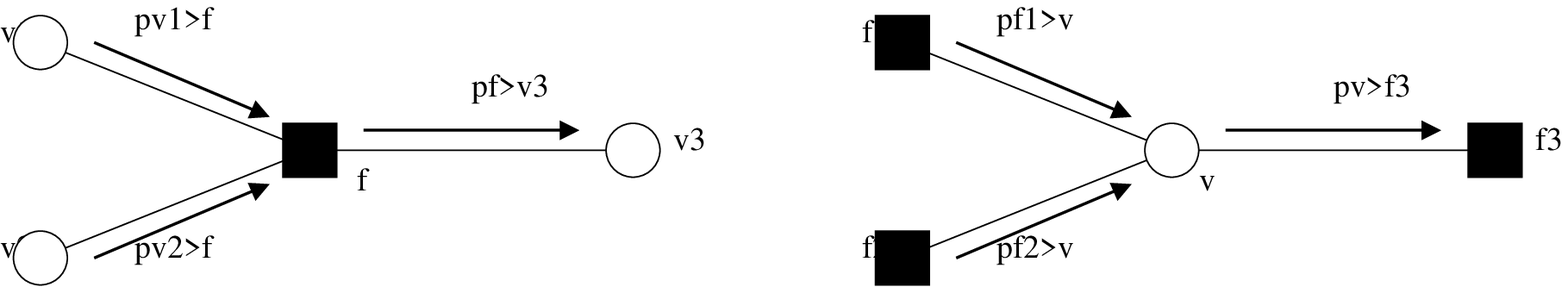}
	{Examples of belief propagation among nodes of a factor graph.}
	{3.25in}
	{\psfrag{v1}[r][Bl][1.0]{$v_1$}
	 \psfrag{v2}[r][Bl][1.0]{$v_2$}
	 \psfrag{v3}[l][Bl][1.0]{$v_3$}
	 \psfrag{f}[][Bl][1.0]{$f$}
	 \psfrag{pv1>f}[Bl][Bl][1.0]{$p_{v_1\rightarrow f}$}
	 \psfrag{pv2>f}[tl][Bl][1.0]{$p_{v_2\rightarrow f}$}
	 \psfrag{pf>v3}[B][Bl][1.0]{$p_{f\rightarrow v_3}$}
	 \psfrag{f1}[r][Bl][1.0]{$f_1$}
	 \psfrag{f2}[r][Bl][1.0]{$f_2$}
	 \psfrag{f3}[l][Bl][1.0]{$f_3$}
	 \psfrag{v}[t][Bl][1.0]{$v$}
	 \psfrag{pf1>v}[Bl][Bl][1.0]{$p_{f_1\rightarrow v}$}
	 \psfrag{pf2>v}[tl][Bl][1.0]{$p_{f_2\rightarrow v}$}
	 \psfrag{pv>f3}[B][Bl][1.0]{$p_{v \rightarrow f_3}$}
	}

When the factor graph contains no loops, SPA-BP yields exact 
posteriors after only two rounds of message passing 
(i.e., forward and backward passes). 
However, with loops in the factor graph, convergence to the exact posteriors 
is not guaranteed, as exact inference is known to be NP-hard 
\cite{Cooper:AI:90}.
That said, there exist many problems to which loopy BP has been 
successfully applied, including 
inference on Markov random fields \cite{Freeman:IJCV:00},
multiuser detection \cite{Boutros:TIT:02,Guo:ISIT:07},
turbo decoding \cite{McEliece:JSAC:98},
LDPC decoding \cite{MacKay:Book:03},
and compressed sensing \cite{Rangan:10v2,Baron:TSP:10,Donoho:PNAS:09}. 
Our work not only leverages these past successes, but unites the 
last two through ``turbo'' message scheduling on a larger factor
graph \cite{Schniter:CISS:10}.

\subsection{Background on RBP}				\label{sec:rbp}

A sub-problem of particular interest to us is the estimation of a
non-Gaussian vector $\vec{x}$ that is linearly mixed to form 
$\vec{z}=\vec{\Phi x}$ and subsequently observed as $\vec{y}$ 
through componentwise non-Gaussian measurements 
$\{p_{Y_i|Z_i}(y_i|z_i)\}_{i=0}^{N-1}$.
In our case \eqref{pxj} specifies the non-Gaussian prior on $\vec{x}$ and
\eqref{yi} yields the non-Gaussian measurement 
(where the non-Gaussianity 
results from the inherent uncertainty on data symbols $s_i$). 
This sub-problem yields the factor graph shown within the 
right dashed box in \figref{factor_graph_noncoh},
where the nodes ``$y_i$'' represent the measurements
and the rightmost nodes represent the prior on $\vec{x}$.

Building on prior multiuser detection work by Guo and Wang 
\cite{Guo:ISIT:07}, Rangan recently proposed a so-called 
``relaxed BP'' (RBP) scheme \cite{Rangan:10v2} that yields
asymptotically exact posteriors as $N,L\rightarrow\infty$ 
(under some additional technical conditions on $\vec{\Phi}$) 
\cite{Rangan:10v2}.
The main ideas behind RBP are the following.
First, although the beliefs flowing leftward from the nodes $\{x_j\}$ 
are clearly non-Gaussian, the corresponding belief about 
$z_i = \sum_{j=0}^{L-1}\Phi_{ij} x_j$ can be accurately approximated as 
Gaussian, when $L$ is large, using the central limit theorem.
Moreover, to calculate the parameters of this distribution (i.e., its
mean and variance), only the mean and variance of each $x_j$ are needed.
Thus, it suffices to pass only means and variances leftward from each
$x_j$ node.
It is similarly desirable to pass only means and variances rightward
from each measurement node.
Although the exact rightward flowing beliefs would be non-Gaussian (due to
the non-Gaussian assumption on the measurement channels $p_{Y_i|Z_i}$),
RBP approximates them as Gaussian using a 2nd-order Taylor series,
and passes only the resulting means and variances.
A further simplification employed by RBP is to approximate
the \emph{differences} among the outgoing means/variances of each left node,
and the incoming means/variances of each right node, using Taylor series.
The RBP algorithm\footnote{
  To be precise, the RBP algorithm in \tabref{rbp} is an extension of 
  that proposed in \cite{Rangan:10v2}.
  \tabref{rbp} handles \emph{complex} Gaussian distributions
  and \emph{non-identically distributed} signal and measurement channels.} 
is summarized in \tabref{rbp}.
Assuming (D1)-(D6) can be calculated efficiently (as is the case in our
problem), the complexity of RBP is $\mc{O}(NL)$.

\putTable{rbp}{The RBP Algorithm}{
\begin{equation*}
\begin{array}{|lrcl@{}r|}\hline
  \multicolumn{2}{|l}{\textsf{definitions:}}&&&\\[-1mm]
  &p_{Z_i|Y_i}(z|y;\hat{z},\mu^z)
   &=& \frac{p_{Y_i|Z_i}(y|z) \,\mc{CN}(z;\hat{z},\mu^z)}
	{\int_{z'} p_{Y_i|Z_i}(y|z') \,\mc{CN}(z';\hat{z},\mu^z)} &\text{(D1)}\\
  &F\out(y,\hat{z},\mu^z)
   &=& \int_z z\, p_{Z_i|Y_i}(z|y;\hat{z},\mu^z) &\text{(D2)}\\
  &\mc{E}\out(y,\hat{z},\mu^z)
   &=& \int_z |z-F\out(y,\hat{z},\mu^z)|^2\, 
   	p_{Z_i|Y_i}(z|y;\hat{z},\mu^z)&\text{(D3)}\\
  &p_{Q_j}\!(q;\hat{q},\mu^q)
   &=& \frac{p_{X_j}\!(q) \,\mc{CN}(q;\hat{q},\mu^q)}
   	{\int_{q'}p_{X_j}\!(q') \,\mc{CN}(q';\hat{q},\mu^q)}&\text{(D4)}\\
  &F\inp(\hat{q},\mu^q)
   &=& \int_q q\, p_{Q_j}\!(q;\hat{q},\mu^q) &\text{(D5)}\\
  &\mc{E}\inp(\hat{q},\mu^q)
   &=& \int_q |q-F\inp(\hat{q},\mu^q)|^2\, p_{Q_j}\!(q;\hat{q},\mu^q) &\text{(D6)}\\
  \multicolumn{2}{|l}{\textsf{initialize:}}&&&\\
  &\forall i,j: 
   \hat{x}_{ij}[1] &=& \hat{x}_{j}[1] 
	\,=\, \int_{x} x\, p_{X_j}(x) & \text{(I1)}\\
  &\forall j:
   \mu^x_j[1] &=& \int_{x} |x-\hat{x}_j[1]|^2  p_{X_j}(x) & \text{(I2)}\\
  \multicolumn{2}{|l}{\textsf{for $n=1,2,3,\dots$}}&&&\\
  &\forall i:
   \mu^z_i[n]
   &=& \textstyle \sum_{j=0}^{L-1} |\Phi_{ij}|^2 \mu^x_{j}[n] & \text{(R1)}\\
  &\forall i:
   \hat{z}_i[n]
   &=& \textstyle \sum_{j=0}^{L-1} \Phi_{ij} \hat{x}_{ij}[n] & \text{(R2)}\\
  &\forall i,j:
   \hat{z}_{ij}[n]
   &=& \hat{z}_i[n] - \Phi_{ij} \hat{x}_{ij}[n] & \text{(R3)}\\
  &\forall i:
   \mu^e_i[n]
   &=& \mc{E}\out(y_i,\hat{z}_i[n],\mu^z_i[n]) & \text{(R4)}\\
  &\forall i,j:
   \hat{e}_{ij}[n]
   &=& F\out(y_i,\hat{z}_i[n],\mu^z_i[n]) - \hat{z}_{ij}[n]
	&\\&&&\mbox{}
	- \Phi_{ij}\hat{x}_{ij}[n]
        \mu^e_i[n]/\mu^z_i[n] & \text{(R5)}\\
  &\forall i:
   \mu^u_i[n]
   &=& \big(1-\mu^e_i[n]/\mu^z_i[n]\big)^{-1} \mu^z_i[n] & \text{(R6)}\\
  &\forall i,j:
   \hat{u}_{ij}[n]
   &=& \big(1-\mu^e_i[n]/\mu^z_i[n]\big)^{-1} \hat{e}_{ij}[n] & \text{(R7)}\\
  &\forall j:
   \mu^q_j[n]
   &=& \textstyle \big(\sum_{i=0}^{N-1} |\Phi_{ij}|^2/\mu^u_i[n] 
	\big)^{-1} & \text{(R8)}\\
  &\forall j:
   \hat{q}_j[n]
   &=& \textstyle \mu^q_j[n] \sum_{i=0}^{N-1} \big( \Phi_{ij}^*
	\hat{u}_{ij}[n]/\mu^u_i[n]\big)  & \text{(R9)}\\
  &\forall j:
   \mu^x_j[n\!+\!1]
   &=& \mc{E}\inp(\hat{q}_j[n],\mu^q_j[n]) & \text{(R10)}\\
  &\forall j:
   \hat{x}_{j}[n\!+\!1]
   &=& F\inp(\hat{q}_j[n],\mu^q_j[n]) & \text{(R11)}\\
  &\forall i,j:
   \hat{x}_{ij}[n\!+\!1]
   &=& \hat{x}_{j}[n\!+\!1] - \big( \Phi_{ij}^* 
	\hat{u}_{ij}[n]/\mu^u_i[n]\big) \mu^x_j[n\!+\!1] & \text{(R12)}\\
  \multicolumn{2}{|l}{\textsf{end}}&&&\\\hline
\end{array}
\end{equation*}
}

\subsection{BP-based joint channel estimation and decoding}	\label{sec:jced}

In this section, we detail our BP-based approach to JCED,
frequently referring to the factor graph in \figref{factor_graph_noncoh}.
Note that, since our factor graph is loopy, there exists considerable 
freedom in the message passing schedule.
We choose to propagate beliefs from the left to the right and back again,
several times, stopping as soon the beliefs have appeared to converge. 
Each full cycle of message passing on the overall factor graph will
be referred to as a ``turbo iteration.''
During each turbo iteration, several rounds of message passing are
performed within each of the dashed boxes in \figref{factor_graph_noncoh}.
We refer to the iterations within the left dashed box as 
``SISO decoder iterations'' and the iterations within the right dashed
as ``RBP iterations.''
Below, we provide details on how beliefs are calculated and propagated.

At the very start, nothing is known about the information bits, which
are assumed apriori to be equally likely
(i.e., $\Pr\{b_m\!=\!1\}\!=\!\frac{1}{2}~\forall m$). 
Thus, the bit beliefs that initially flow rightward
out of the coding/interleaving block are uniform
(i.e., $p_{c_{i,m}\rightarrow \mc{M}_i}(1)=\frac{1}{2}$ for all
indices $(i,m)$ corresponding to data bits).
Meanwhile, the values of the pilot/training bits are known with certainty, 
and so $p_{c_{i,m}\rightarrow \mc{M}_i}(c)=1$ for $c=c_{i,m}$.

Next, coded-bit beliefs are propagated rightward into the symbol
mapping nodes. 
Since the symbol mapping is deterministic, the pdf factors take the form 
$p(s\of{k}\giv\vec{c}\of{l}) = \delta_{k-l}$, where $\{\delta_k\}_{k\in\Int}$ 
denotes the Kronecker delta sequence.
According to the SPA, the message passed 
rightward from symbol mapping node ``$\mc{M}_i$'' takes the form
\begin{eqnarray}
  p_{\mc{M}_i\rightarrow s_i}(s\of{k})
  &\propto& \sum_{\vec{c}\in\{0,1\}^M} p(s\of{k}\giv\vec{c}) \prod_{m=1}^M 
  	p_{c_{i,m}\rightarrow \mc{M}_i}(c_m) \quad \\
  &=& \prod_{m=1}^M p_{c_{i,m}\rightarrow \mc{M}_i}(c_m\of{k}) .
\end{eqnarray}
The SPA then implies that the same message is passed rightward from node $s_i$ 
(i.e., $p_{\mc{M}_i\rightarrow s_i}(s\of{k})=p_{s_i\rightarrow y_i}(s\of{k})$).

Recall, from the discussion of RBP, that the belief propagating rightward
into the OFDM observation node ``$y_i$'' determines RBP's $i^{th}$ 
measurement pdf $p_{Y_i|Z_i}(y|z)$.
Writing this belief as $\beta_i\of{k}\defn p_{s_i\rightarrow y_i}(s\of{k})$,
\eqref{yi} implies a Gaussian-mixture channel of the form 
\begin{eqnarray}
  p_{Y_i|Z_i}(y|z)
  &=& \sum_{k=1}^{2^M} \beta_i\of{k} \,\mc{CN}(y;s\of{k}z,\mu^v) ,
  							\label{eq:pY|Z}
\end{eqnarray}
From \eqref{pY|Z}, it can be shown (see \appref{out}) that the 
quantities (D2)-(D3) in \tabref{rbp} become
\begin{eqnarray}
  F\out(y,\hat{z},\mu^z)
  &=& \hat{z} + \hat{e}_i(y,\hat{z},\mu^z) 		\label{eq:Fout}\\
  \mc{E}\out(y,\hat{z},\mu^z)
  &=& \sum_{k=1}^{2^M} \xi_i\of{k}(y,\hat{z},\mu^z) \Big( 
	\frac{\mu^z\mu^v}{|s\of{k}|^2 \mu^z+\mu^v} 
	+ \big|\hat{e}_i(y,\hat{z},\mu^z)-\hat{e}\of{k}(y,\hat{z},\mu^z)\big|^2
	\Big) 						\label{eq:Eout}
\end{eqnarray}
for
\begin{eqnarray}
  \hat{e}\of{k}(y,\hat{z},\mu^z)
  &\defn& \Big(\frac{y}{s\of{k}}-\hat{z}\Big) 
  	\frac{|s\of{k}|^2\mu^z}{|s\of{k}|^2\mu^z+\mu^v} \label{eq:ek}\\
  \xi_i\of{k}(y,\hat{z},\mu^z)
  &\defn& \frac{\beta_i\of{k}
  		\mc{CN}(y;s\of{k}\hat{z},|s\of{k}|^2\mu^z\!+\!\mu^v)}
           {\sum_{k'} \beta_i\of{k'}
	    	\mc{CN}(y;s\of{k'}\hat{z},|s\of{k'}|^2\mu^z\!+\!\mu^v)} 
							\label{eq:xi}\\
  \hat{e}_i(y,\hat{z},\mu^z)
  &\defn& \sum_{k=1}^{2^M} \xi_i\of{k}\!(y,\hat{z},\mu^z)
  	\,\hat{e}\of{k}(y,\hat{z},\mu^z) .		\label{eq:ei}
\end{eqnarray}
The quantities in \eqref{Fout}-\eqref{Eout} can be interpreted as follows.
Given the observation $y_i=y$ and assuming the prior 
$z_i\sim\mc{CN}(\hat{z},\mu^z)$ on the subcarrier gain $z_i$, the quantity
$F\out(y,\hat{z},\mu^z)$ is the MMSE estimate of $z_i$,
$\mc{E}\out(y,\hat{z},\mu^z)$ is its variance,
and $\{\xi_i\of{k}(y,\hat{z},\mu^z)\}_{k=1}^{2^M}$ 
is the posterior pmf of $s_i$.
Likewise, from \eqref{pxj}, it can be shown (see \appref{in})
that the quantities (D5)-(D6) in \tabref{rbp} take the form 
\begin{eqnarray}
  F\inp(\hat{q},\mu^q)
  &=& \frac{\gamma_j(\hat{q},\mu^q)}{\alpha_j(\hat{q},\mu^q)} 
  							\label{eq:Fin}\\
  \mc{E}\inp(\hat{q},\mu^q)
  &=& |\gamma_j(\hat{q},\mu^q)|^2 
	\frac{\alpha_j(\hat{q},\mu^q)-1}{[\alpha_j(\hat{q},\mu^q)]^2} 
	+ \frac{\nu_j(\mu^q)}{\alpha_j(\hat{q},\mu^q)},	\label{eq:Ein} 
\end{eqnarray}
for
\begin{eqnarray}
   \alpha_j(\hat{q},\mu^q)
   &\defn& 1 + \frac{1-\lambda_j}{\lambda_j} 
	\frac{\mu_j}{\nu_j(\mu^q)} 
        \exp\!\left(
	-\frac{|\gamma_j(\hat{q},\mu^q)|^2}{\nu_j(\mu^q)} 
	\right) 					\label{eq:alf}\\
   \gamma_j(\hat{q},\mu^q)
   &\defn& \frac{\nu_j(\mu^q)}{\mu^q}\hat{q} 		\label{eq:gam}\\
   \nu_j(\mu^q)
   &\defn& \frac{\mu^q\mu_j}{\mu^q + \mu_j} .		\label{eq:nu}
\end{eqnarray}
The quantity $F\inp$ from \eqref{Fin} can be interpreted as the MMSE 
estimate of the channel tap $x_j$ given the observations $\vec{y}$
and the pilots $\vec{c}\pt$, and the quantity $\mc{E}\inp$ from \eqref{Ein} 
can be interpreted as its variance.

Using the quantities derived in \eqref{Fout}-\eqref{nu}, the RBP algorithm 
in \tabref{rbp} is iterated until convergence is detected.
Doing so generates approximately conditional-mean 
(i.e., nonlinear MMSE) estimates $\{\hat{x}_j\}$ of the sparse-channel 
impulse-response coefficients $\{x_j\}$, 
as well as their conditional variances $\{\mu_j^x\}$,
based on the observations $\{y_i\}$ and the soft symbol estimates
$\{\beta_{i}\of{k}\}$. 
Conveniently, RBP also returns (a close approximation to)
the conditional-mean estimates $\{\hat{z}_i\}$ of the subchannel gains 
$\{z_i\}$, as well as their conditional variances $\{\mu_i^z\}$.

Before continuing, we discuss some RBP details that are specific to our
JCED application.
First, we notice that the condition $\mu_i^e[n]\!<\!\mu_i^z[n]$ is required to
guarantee a positive value of the variance $\mu_i^u[n]$ in (R6). 
Intuitively, we might expect that $\mu_i^e[n]\!<\!\mu_i^z[n]$, because
$\mu_i^e[n] \!=\! \mc{E}\out(y_i,\hat{z}_i[n],\mu_i^z[n])$ is a posterior 
variance and $\mu_i^z[n]$ a prior variance.
However, this is not necessarily the case during the first few RBP iterations, 
when the soft channel and symbol estimates may be inaccurate.
We remedy this situation by clipping $\mu^e_i[n]$ at the value $0.99\mu^z[n]$, 
where $0.99$ was chosen heuristically.
Second, due to the DFT matrix property $|\Phi_{ij}|^2\!=\!1~\forall i,j$,
step (R1) in \tabref{rbp} simplifies to 
$\mu_i^z[n] \!=\! \mu^z[n] \!\defn\! \sum_{j=0}^{L-1} \mu_j^x[n]$,
and (R8) simplifies to
$\mu_j^q[n] \!=\! \mu^q[n] \!\defn\! \big(\sum_{i=0}^{N-1} 1/\mu_i^u[n]\big)^{-1}$.
With these simplifications, the complexity of RBP is dominated by the
computation of the elementwise matrix products $\Phi_{ij}\hat{x}_{ij}$
and $\Phi_{ij}^*\hat{u}_{ij}$, which must each be calculated once per
RBP iteration, as well as three other elementwise matrix products in 
(R5), (R7), and (R12).
Thus, RBP requires only $\approx 5NL$ multiplies per iteration.

After RBP converges, updated symbol beliefs are passed leftward out of 
the RBP sub-graph.
According to the SPA, the belief propagating leftward 
from the $y_i$ node takes the form 
\begin{eqnarray}
  p_{s_i\leftarrow y_i}(s)
  &\propto& \int_{z} \mc{CN}(y_i;s z,\mu^v) \,\mc{CN}(z;\hat{z}_i,\mu_i^z) \\
  &=& \mc{CN}(y_i;s \hat{z}_i,|s|^2\mu_i^z+\mu^v) ,
\end{eqnarray}
where the quantities $(\hat{z}_i,\mu_i^z)$ play the role of
soft channel estimates. 
The SPA then implies that 
$p_{\mc{M}_i\leftarrow s_i}(s)=p_{s_i\leftarrow y_i}(s)$.

Next, beliefs are passed leftward from each symbol-mapping node $\mc{M}_i$ 
to the corresponding bit nodes $c_{i,m}$. 
From the SPA, these beliefs take the form
\begin{eqnarray}
  p_{c_{i,m}\leftarrow\mc{M}_i}(c) 
  &\propto& \sum_{k=1}^{2^M} \sum_{\vec{c}:c_m=c} p(s\of{k}\giv\vec{c}) 
  	~ p_{\mc{M}_i\leftarrow s_i}(s\of{k}) 
	\prod_{m'\neq m} p_{c_{i,m'}\rightarrow\mc{M}_i}(c_{m'}) \nonumber\\
  &=& \sum_{k: c_m\of{k}=c} p_{\mc{M}_i\leftarrow s_i}(s\of{k})
  	\frac{\prod_{m'=1}^M p_{c_{i,m'}\rightarrow\mc{M}_i}(c_{m'}\of{k})}
	{p_{c_{i,m}\rightarrow\mc{M}_i}(c)} \\
  &=& \frac{1}{p_{c_{i,m}\rightarrow\mc{M}_i}(c)} 
  	\sum_{k: c_m\of{k}=c} p_{\mc{M}_i\leftarrow s_i}(s\of{k})
  	p_{\mc{M}_i\rightarrow s_i}(s\of{k})  
\end{eqnarray}
for pairs $(i,m)$ that do not correspond to pilot/training bits.
(Since the pilot/training bits are known with certainty, there is no
need to update their pmfs.)

Finally, messages are passed leftward into the coding/interleaving block.
Doing so is equivalent to feeding extrinsic soft bit estimates to
a soft-input/soft-output (SISO) deinterleaver/decoder,
which treats them as priors.
Since SISO decoding is a well-studied topic (see, e.g., 
\cite{MacKay:Book:03,Richardson:Book:09}) and high-performance implementations 
are readily available, we will not elaborate on the details here.
It suffices to say that, once the extrinsic outputs of the SISO decoder 
have been computed, they are re-interleaved and passed rightward from the 
code/interleave block to begin another round of belief propagation on the 
overall factor graph of \figref{factor_graph_noncoh}. 
The outer ``turbo'' iterations then continue until either the 
decoder detects no bit errors, the soft bit estimates have converged, or a
maximum number of iterations has elapsed.

\section{Numerical Results}				\label{sec:sims}

In this section, we present numerical results that compare our 
proposed BP-JCED to the CCS approach as well as to
several reference schemes that act as performance upper/lower bounds.

\subsection{Setup and reference schemes}

The following decoupled channel-estimation and decoding (DCED) procedure 
was used to implement CCS.
First, a LASSO\footnote{
  The criterion employed by LASSO \cite{Tibshirani:JRSSb:96} is equivalent 
  to the one employed in ``basis pursuit denoising'' \cite{Chen:JSC:98}.}
channel estimate $\hvec{x}[t]$ was generated using pilot-subcarriers.
To implement LASSO, we used the celebrated SPGL1 algorithm 
\cite{vandenBerg:JSC:08} with a genie-optimized tuning parameter.\footnote{
  The performance of LASSO/SPGL1 is highly dependent on the value of a 
  tuning parameter that determines the tradeoff between the estimate's 
  sparsity and the residual's variance.
  To optimize this tradeoff, for each realization, SPGL1 was invoked over 
  a dense grid of tuning parameters, and the one that minimized $\NMSE$
  (with respect to the true channel) was chosen.}
The frequency-domain estimate $\hvec{z}[t]\!=\!\vec{\Phi}\hvec{x}[t]$
was then computed, from which the (genie-aided empirical) variance 
$\hat{\mu}^z[t]\defn\norm{\hvec{z}[t]-\vec{z}[t]}_2^2/N$ was calculated.
Using the soft channel estimate $(\hvec{z}[t],\hat{\mu}^z)$, 
leftward SPA-BP on the factor graph in \figref{factor_graph_noncoh}
was performed exactly as described in \secref{jced}, ensuring that the 
LASSO outputs were properly combined with SISO decoding.
We note that, due to the two genie-aided steps, the performance attained 
by CCS may be somewhat optimistic.
Even so, we shall see that this optimistic CCS performance remains far
below that of our BP-based JCED approach (which requires no genie-aided
steps).

We now describe several reference schemes, all of which use the DCED 
procedure described above, but with different channel estimators.
The first uses traditional linear MMSE (LMMSE) estimation.
Since LMMSE does not exploit channel sparsity, it yields a performance
lower-bound for any sparsity-leveraging technique. 
We also consider MMSE-optimal\footnote{
  When the sparse-channel support is known, the non-zero channel 
  coefficients follow a Gaussian prior, and MMSE-optimal estimates
  can be calculated linearly.}
pilot-aided channel estimation under the 
\emph{support-aware genie} (SG), reasoning that this yields a performance 
upper-bound for CCS.
Finally, we consider MMSE-optimal estimation under a 
\emph{bit- and support-aware genie} (BSG).
Here, in addition to the channel support being known, all bits 
(including data bits) are known and used for channel estimation.
This latter reference scheme yields a performance upper-bound for 
\emph{any} implementable DCED or JCED scheme, including our BP-based JCED.
Remarkably, we shall see that that performance of our proposed scheme
is not far from that of the BSG.

For all of our results, we used
irregular LDPC codes with codeword length $\approx\! 10000$ and 
average column weight $3$,
generated (and decoded) using the publicly available software 
\cite{Kozintsev:SW}.
Random interleaving did not seem to have an effect, and so no interleaving
was employed.
For bit-to-symbol mapping, we used multilevel Gray-mapping \cite{deJong:TCOM:05}, 
noting recent work \cite{Samuel:ASIL:09} that conjectures the optimality 
of Gray-mapping when BICM is used with a strong code.
For OFDM, we used\footnote{
  Experiments with non-prime $N\!=\!1024$ showed a slight
  degradation of performance.} 
$N\!=\!1021$ subcarriers, since prime $N$ 
ensures that square/tall submatrices of $\vec{\Phi}$ will be full-rank.
As described in the sequel, we tested various combinations of
pilot subcarriers $\Np$ and interspersed training bits $\Mt$.
The $\Np$ pilot subcarriers were spaced uniformly and modulated
with QAM symbols chosen uniformly at random.
The $\Mt$ training bits were placed at the most significant bits (MSBs) 
of uniformly spaced data subcarriers with values chosen uniformly at random.

Unless otherwise specified, we used length $L\!=\!256$ channels with 
sparsity rate $\lambda\!=\!1/4$, yielding 
$\E\{\Ls\}\!=\!\lambda N\!=\!64$ non-zero taps on average.
All results are averaged over $T\!=\!100$ OFDM symbols.

\subsection{$\NMSE$ and $\BER$ versus the number of pilot subcarriers}

\putFrag{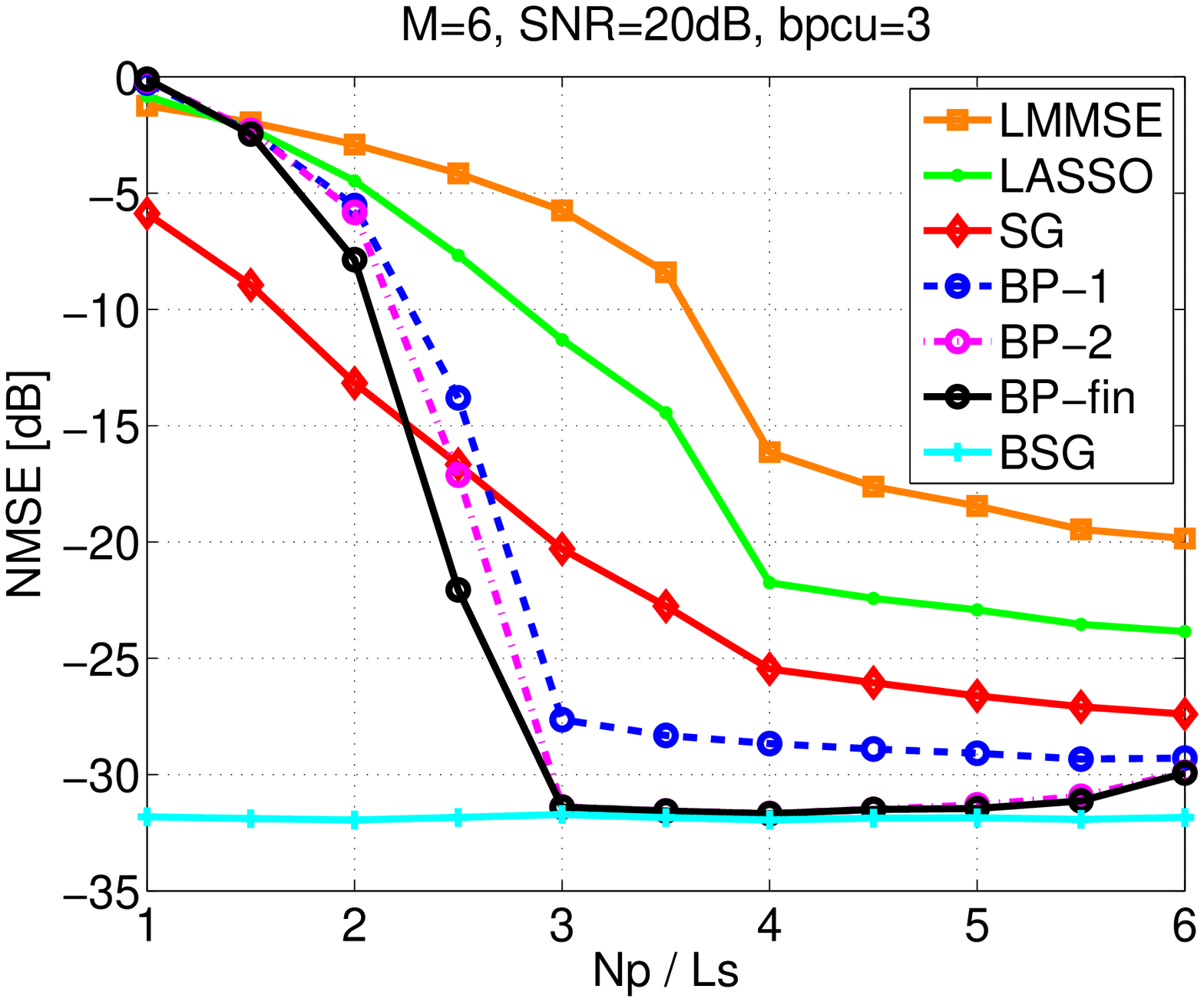}
	{Channel estimation $\NMSE$ versus pilot-to-sparsity ratio $\Np/\Ls$, 
	 for
	 $\SNR\!=\!20$dB, 
	 $\Mt\!=\!0$ training bits, 
	 $\eta\!=\!3$ bpcu, 
	 and 64-QAM.}
	{\figsize}
	{\psfrag{Np / Ls}[][][0.9]{$\Np/\Ls$} 
	 \psfrag{LASSO}[Bl][Bl][0.87]{\hspace{-0.3mm}\sf CCS} 
	 \psfrag{BP-fin}[Bl][Bl][0.87]{\hspace{-0.3mm}\sf BP--$\infty$} 
	 \psfrag{NMSE [dB]}[][][0.9]{\sf $\NMSE$ [dB]}
	 \psfrag{M=6, SNR=20dB, bpcu=3}[][][0.9]{}}

\Figref{mse_vs_Np_20} plots channel estimation normalized mean-squared
error $\NMSE$
$\!\defn\!\norm{\hvec{x}[t]-\vec{x}[t]}^2_2/\norm{\vec{x}[t]}^2_2$
versus the pilot-to-sparsity ratio $\Np/\Ls$ at $\SNR\!=\!20$dB.
As expected, CCS's $\NMSE$ falls between that of LMMSE and SG estimators, 
and all three decrease monotonically with $\Np/\Ls$.
Even after a single turbo iteration, BP-JCED significantly outperforms
CCS, and---perhaps surprisingly---the SG (when $\Np/\Ls\!\geq\! 3$).
The reason for this latter behavior is that, while the SG uses only the $\Np$ 
pilot subcarriers, BP-JCED uses all $N$ subcarriers, which yields improved
performance even though the $\Nd\!=\!N\!-\!\Np$ data symbols are 
known with very little certainty during the first turbo iteration. 
\Figref{mse_vs_Np_20} indicates that, after only 2 turbo iterations, 
BP-JCED learns the data symbols well enough to estimate the channel 
nearly as well as the BSG (which knows the data symbols perfectly).
The fact that BP-JCED can generate channel estimates that are nearly 
as good as BSG's support-aware estimates attests to the near-optimal 
compressive estimation abilities of RBP.

\putFrag{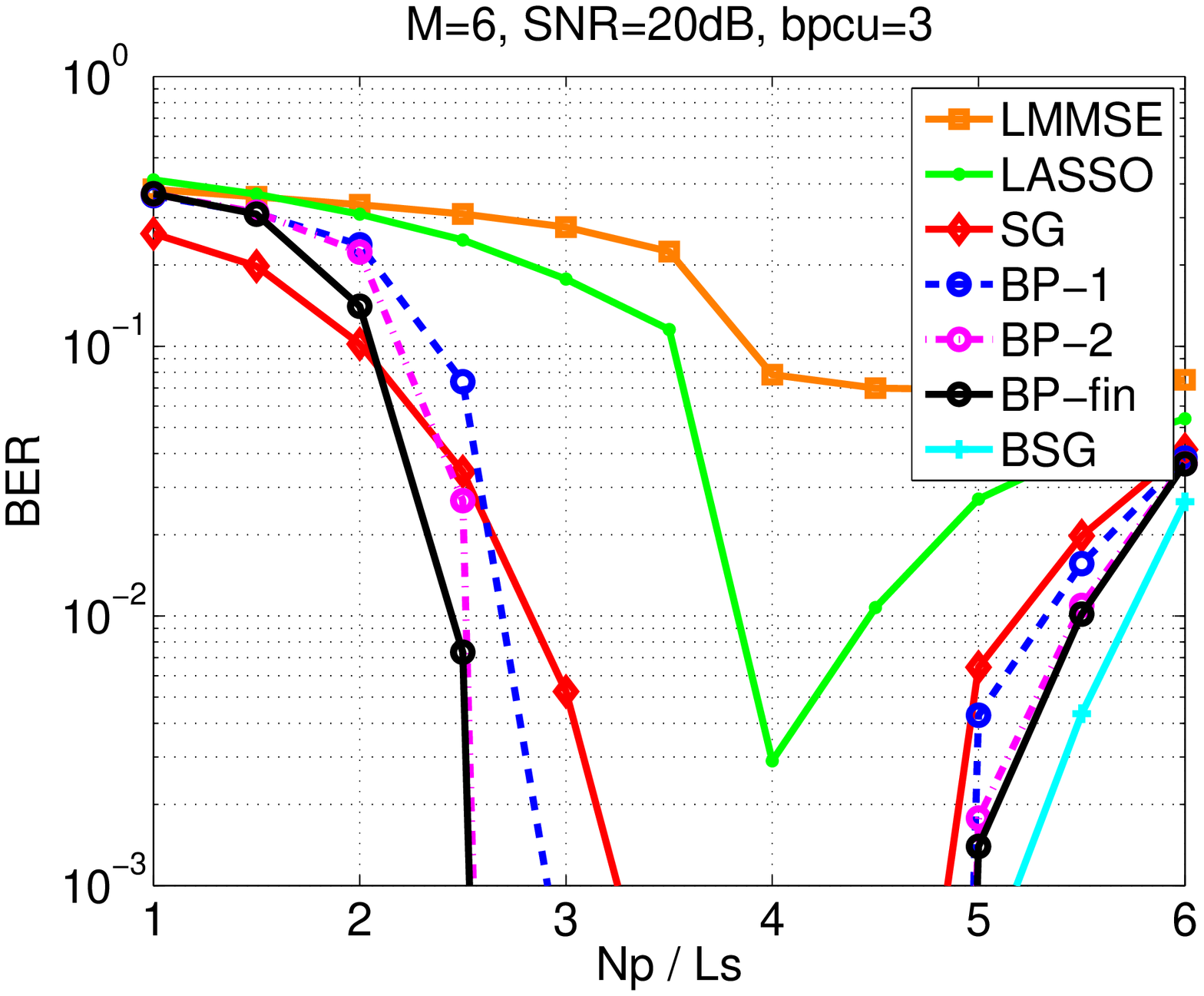}
	{$\BER$ versus pilot ratio $\Np/\Ls$, for
	 $\SNR\!=\!20$dB, 
	 $\Mt\!=\!0$ training bits, 
	 $\eta\!=\!3$ bpcu, 
	 and 64-QAM.}
	{\figsize}
	{\psfrag{Np / Ls}[t][t][0.9]{$\Np/\Ls$} 
	 \psfrag{LASSO}[Bl][Bl][0.87]{\hspace{-0.3mm}\sf CCS} 
	 \psfrag{BP-fin}[Bl][Bl][0.87]{\hspace{-0.3mm}\sf BP--$\infty$} 
	 \psfrag{BER}[][][0.9]{$\BER$}
	 \psfrag{M=6, SNR=20dB, bpcu=3}[][][0.9]{}}

\Figref{ber_vs_Np_20} plots bit error rate ($\BER$) versus the pilot ratio 
$\Np/\Ls$ at $\SNR\!=\!20$dB and a fixed spectral efficiency of 
$\eta\!=\!3$ bpcu.
The curves exhibit a ``notched'' shape because, 
as $\Np$ increases, the code rate $R$ must decrease to maintain a
fixed value of spectral efficiency $\eta$.  
Thus, while an increase in $\Np$ can make channel estimation easier, 
the reduction in $R$ makes data decoding more difficult.
For CCS, \figref{ber_vs_Np_20} indicates that $\Np\!=\!4\Ls\!=\!L$ is optimal.
The SG and BP-JCED curves show a similar notch-like shape, although their 
notches are much wider.
Finally, the degredation of BP-JCED's data-bit estimates at large 
$\Np/\Ls$ explains the degredation of its channel estimates, as seen
in \figref{mse_vs_Np_20}, since, with JCED, channel estimation 
is data-directed.

\putFrag{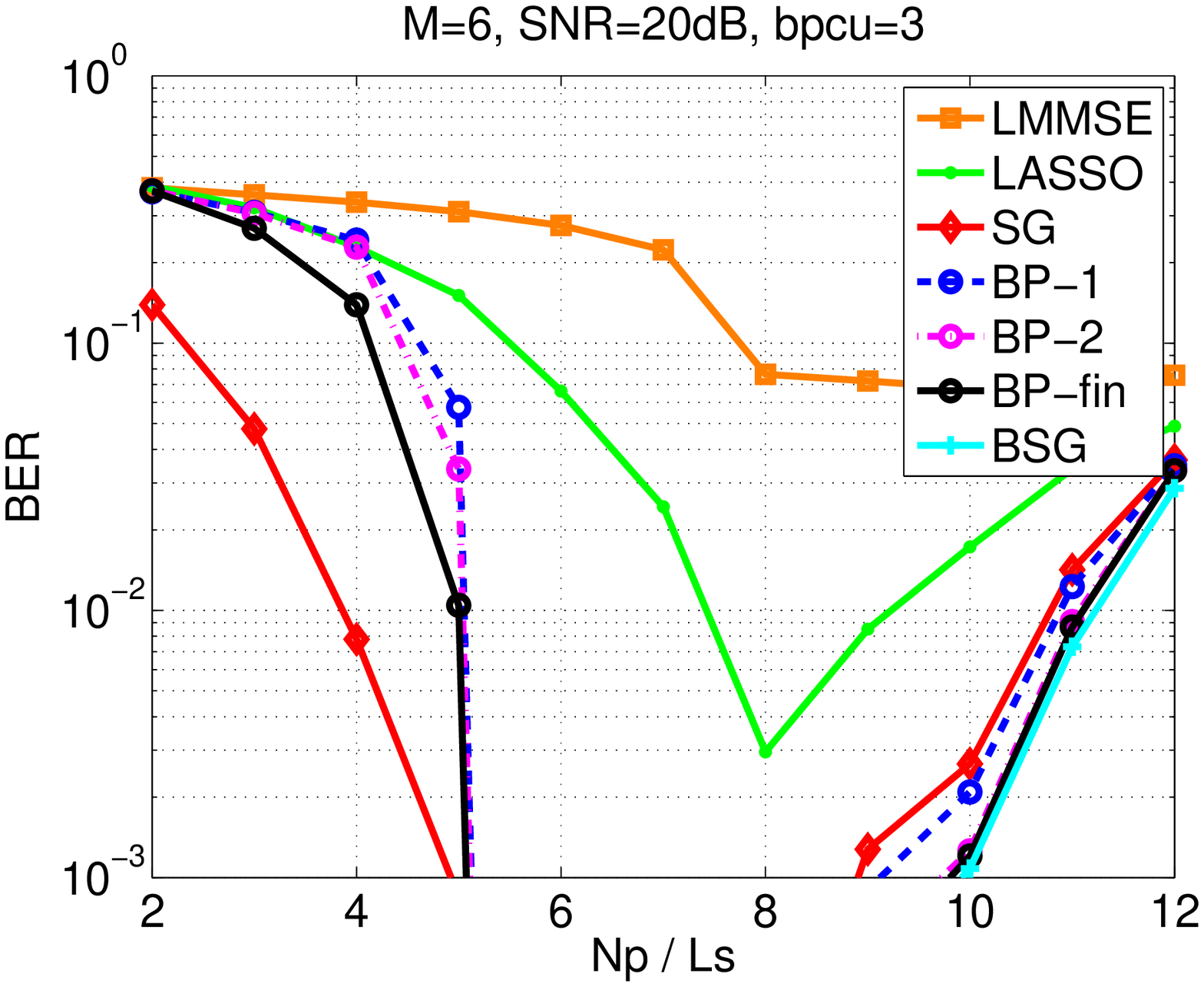}
	{$\BER$ versus pilot ratio $\Np/\Ls$, for
	 $\SNR\!=\!20$dB, 
	 $\Mt\!=\!0$ training bits, 
	 $\eta\!=\!3$ bpcu, 
	 and 64-QAM.
	 The channel used here had the sparsity rate 
	 $\lambda\!=\!\E\{\Ls\}/L\!=\!1/8$, which is
	 half the value used in all other experiments.}
	{\figsize}
	{\psfrag{Np / Ls}[t][t][0.9]{$\Np/\Ls$} 
	 \psfrag{LASSO}[Bl][Bl][0.87]{\hspace{-0.3mm}\sf CCS} 
	 \psfrag{BP-fin}[Bl][Bl][0.87]{\hspace{-0.3mm}\sf BP--$\infty$} 
	 \psfrag{BER}[][][0.9]{$\BER$}
	 \psfrag{M=6, SNR=20dB, bpcu=3}[][][0.9]{}}

It is interesting to notice that \figref{ber_vs_Np_20} shows the optimal 
CCS pilot insertion rate to be the ``Nyquist'' rate of $\Np\!=\!L$,
since, at this pilot rate, CCS is not actually ``compressed.''
To further investigate this behavior, we repeated the experiment 
using a channel with half the number of active coefficients 
(i.e., $\lambda\!=\!\E\{\Ls\}/L\!=\!1/8$) and report the results 
in \figref{ber_vs_Np_20_sparser}.
Remarkably, we find the same behavior: CCS again performs best 
when pilots are inserted at the Nyquist rate of $\Np\!=\!L$.
In fact, we repeated this experiment with dozens of other arbitrary 
combinations of $(N,L,\lambda,\SNR,\eta,M)$, and always found exactly 
the same behavior. 
Our empirical evidence suggests that, generally speaking,
\emph{decoupled sparse-channel estimation and data decoding works best 
when pilots are inserted at the Nyquist rate}, at least for
OFDM signaling under uniform subcarrier power allocation.\footnote{
  It would be interesting to see if this behavior persists when
  the pilot- versus data-subcarrier power allocation is optimized.
  Such an optimization, however, remains outside the scope of this manuscript.}

\subsection{Outage rate and the importance of bit-level training}

\putFrag{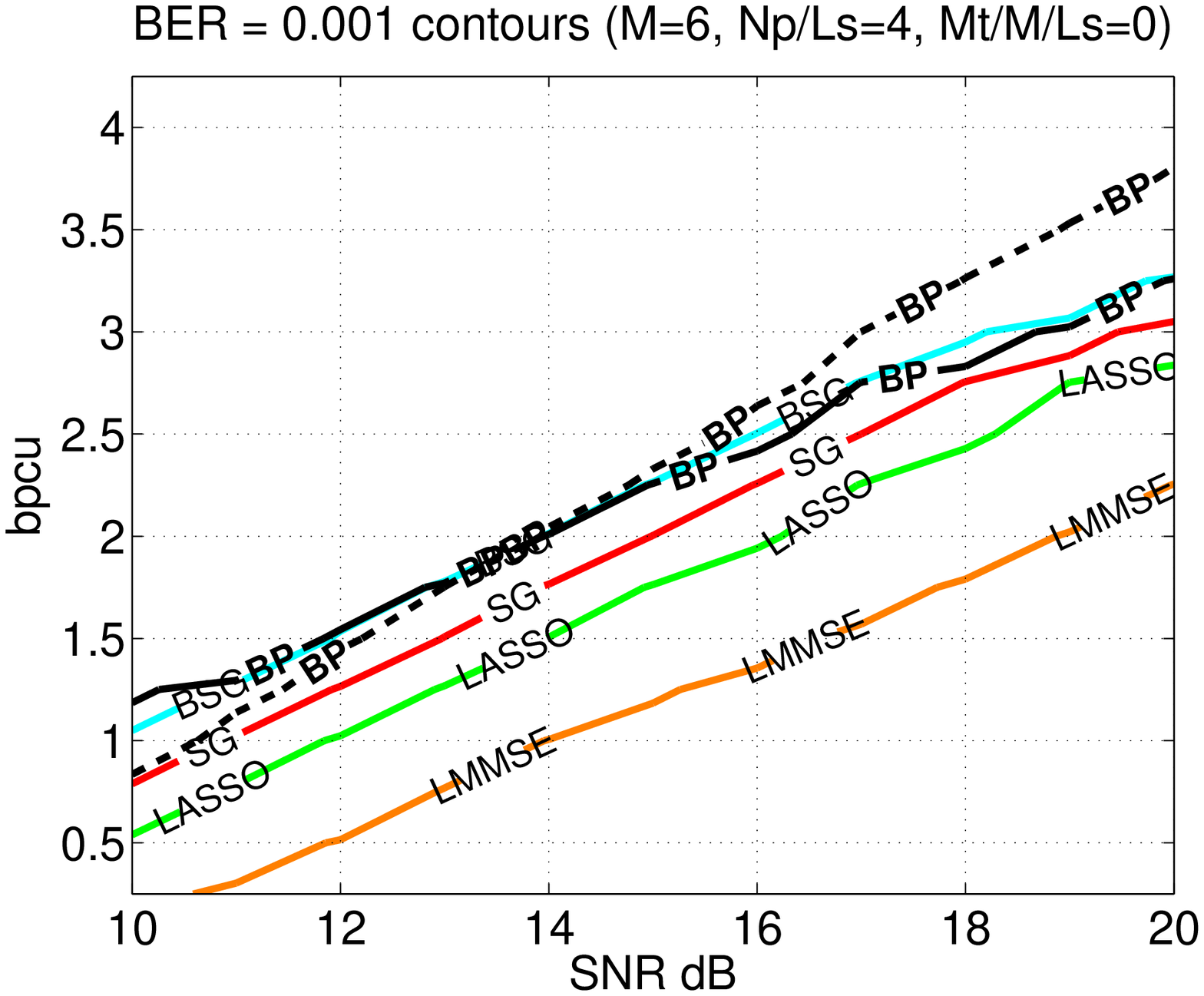}
	{$\BER\!=\!0.001$-achieving spectral efficiency $\eta_{0.001}$ 
	 versus $\SNR$. 
	 The solid traces used $\Np/\Ls\!=\!4$, $\Mt\!=\!0$, and 64-QAM,
	 while the dashed trace used $\Np\!=\!0$, $\Mt\!=\!M\Ls$, and 256-QAM.}
	{\figsize}
	{\psfrag{BER = 0.001 contours (M=6, Np/Ls=4, Mt/M/Ls=0)}[][][0.9]{} 
	 \psfrag{SNR dB}[t][t][0.9]{\sf SNR [dB]} 
	 \psfrag{bpcu}[][][0.9]{\sf bpcu}
	 \psfrag{BSG}[Bl][Bl][0.6]{\sf \textc{BSG}} 
	 \psfrag{SG}[Bl][Bl][0.6]{\sf \textr{SG}} 
	 \psfrag{LASSO}[Bl][Bl][0.6]{\sf \textg{~~\;CCS}} 
	 \psfrag{LMMSE}[Bl][Bl][0.6]{\sf \texto{LMMSE}} }

\Figref{bpcu_vs_snr_combined} plots $\eta_{0.001}$ versus $\SNR$, where
$\eta_{0.001}$ denotes the spectral efficiency (in bpcu) yielding 
$\BER \!=\!0.001$.
The solid-line traces correspond to $\Np\!=\!4\Ls\!=\!L$ pilots,
$\Mt=0$ training bits, and 64-QAM, as suggested by \figref{ber_vs_Np_20}.
These solid-line traces all display the anticipated high-SNR scaling law
$(1-\Np/N)\log_2(\SNR)+\mc{O}(1)$, differing only in the $\mc{O}(1)$ 
offset term.
While, for this setup, we are glad to see BP-JCED performing on par with BSG, 
neither attains the desired channel-capacity prelog-factor of 
$(1-\Ls/N)\!=\!15/16$.
It turns out that this shortcoming is due to the choice $(\Np,\Mt)=(L,0)$, 
which was chosen on behalf of CCS (and not BP-JCED).

To find the optimal choice of $(\Np,\Mt)$ for BP-JCED, we constructed
the $\BER$ plot \figref{Mt_vs_Np_8}. 
There we see that BP-JCED performs best with 
$(\Np,\Mt)\!=\!(0,M\Ls)$, at least in the high-$\SNR$ regime.
Note that the total number of pilot/training bits used when 
$(\Np,\Mt)\!=\!(0,M\Ls)$ is equivalent to $\Ls$ degrees-of-freedom
per fading block, consistent with the channel-capacity prelog factor.
We then evaluated the outage rate of this scheme (with $256$-QAM),
obtaining the dashed $\eta_{0.001}$-vs-$\SNR$ trace in 
\figref{bpcu_vs_snr_combined}, 
which---remarkably---exhibits the desired prelog-factor of $(1-\Ls/N)$.

\putFrag{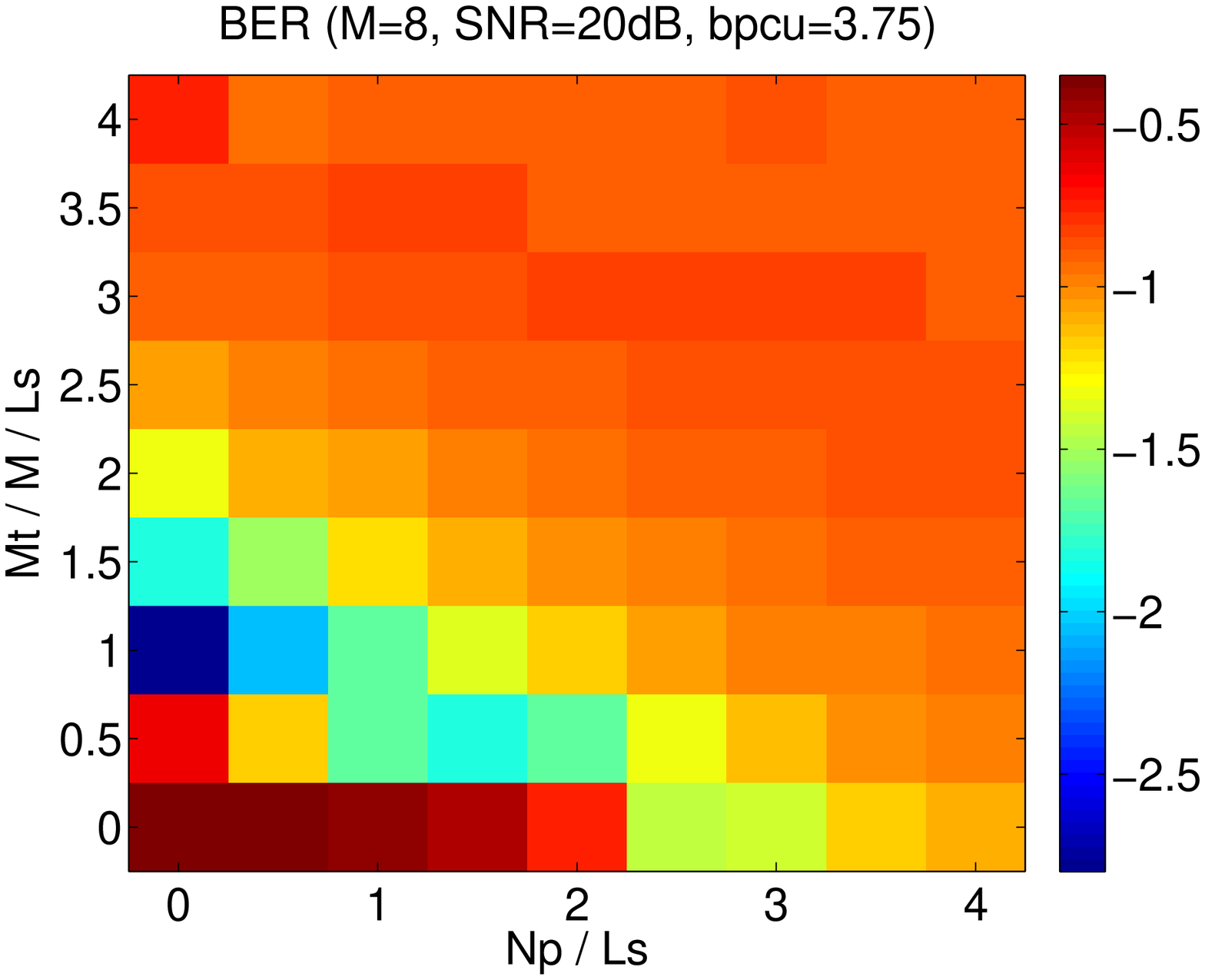}
	{$\log_{10}(\BER)$ versus various combinations of pilot and training rate, for
	 $\SNR\!=\!20$dB, 
	 $\eta\!=\!3.75$ bpcu, 
	 and 256-QAM.}
	{\figsize}
	{\psfrag{Np / Ls}[t][t][0.9]{\sf $\Np/\Ls$} 
	 \psfrag{Mt / M / Ls}[b][b][0.9]{\sf $\Mt/M/\Ls$} 
	 \psfrag{BER (M=8, SNR=20dB, bpcu=3.75)}[][][0.9]{}}

\putFrag{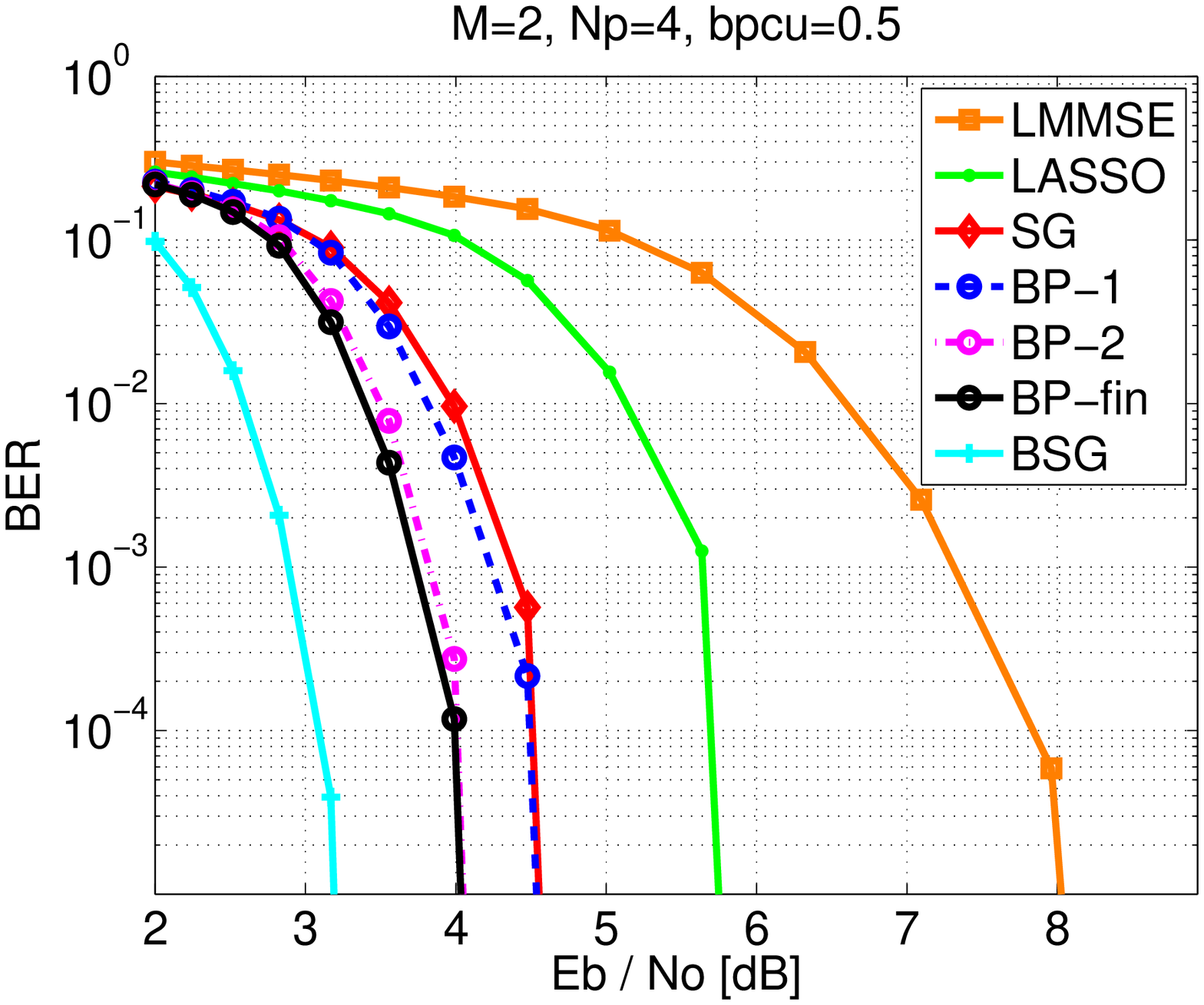}
	{$\BER$ versus $E_b/N_o (\!\defn\!\SNR/\eta)$, for
	 $\Np/\Ls\!=\!4$, $\Mt\!=\!0$, $\eta\!=\!0.5$ bpcu, and 4-QAM.}
	{\figsize}
	{\psfrag{Eb / No [dB]}[t][t][0.9]{\sf $E_b/N_o$ [dB]} 
	 \psfrag{LASSO}[Bl][Bl][0.87]{\hspace{-0.3mm}\sf CCS} 
	 \psfrag{BP-fin}[Bl][Bl][0.87]{\hspace{-0.3mm}\sf BP--$\infty$} 
	 \psfrag{M=2, Np=4, bpcu=0.5}[][][0.9]{}
	 \psfrag{BER}[][][0.9]{$\BER$}}

\Figref{ber_vs_lowsnr} plots $\BER$ versus $E_b/N_o \!\defn\!\SNR/\eta$
over a much lower range of $\SNR$. 
As stated earlier, experiments confirmed that CCS favors $(\Np,\Mt)\!=\!(L,0)$ 
in the low-$\SNR$ regime, and so this configuration was used to keep 
CCS competitive, while being potentially suboptimal for BP-JCED. 
Still, we see from \figref{ber_vs_lowsnr} that BP-JCED, after only two turbo
iterations, beats CCS by $1.8$ dB and remains only $0.8$ dB away from 
the BSG.

\section{Conclusion}				\label{sec:conc}

In this work, we presented a novel approach to joint channel estimation
and decoding (JCED) for spectrally efficient communication over channels 
with possibly sparse impulse responses.
For this, we assumed a pilot-aided transmission scheme that combines bit 
interleaved coded modulation (BICM) with orthogonal frequency division 
multiplexing (OFDM).
Our JCED scheme is based on belief propagation (BP) over a loopy factor graph,
where our BP implementation uses very efficient approximations of the 
sum-product algorithm recently proposed under the guise of 
relaxed belief propagation (RBP) \cite{Guo:ISIT:07,Rangan:10b} and 
soft-input soft-output decoding.
Because our JCED scheme requires only $\approx\! 5NL$ multiplications per 
RBP iteration, we can handle long impulse responses, large numbers of
OFDM subcarriers, and large constellations.
Numerical experiments conducted using $N\!=\!1021$ subcarriers, 
up to $256$-point QAM constellations, $\approx\!10000$-bit LDPC codes,
and channels with length $L=256$ and average sparsity $\E\{\Ls\}=64$,
showed that the $\BER$ of BP-JCED is close to genie-aided bounds and 
much better than the $\BER$ of the LASSO-based ``compressed channel
sensing'' (CCS) approach, where sparse channel estimation is decoupled
from data decoding. 
Moreover, the outage rates observed for BP-JCED exhibit the sparse-channel
capacity pre-log factor $(1\!-\!\Ls/N)$, which is impossible to reach
using CCS.


\appendices

\section{Derivation of RBP Quantities \textnormal{$F\out$} and \textnormal{$E\out$}}	
\label{app:out}

In this appendix, we derive the RBP quantities $F\out(y,\hat{z},\mu^z)$ 
and $\mc{E}\out(y,\hat{z},\mu^z)$ given in \eqref{Fout}-\eqref{ei}.

From (D1)-(D2), we have that 
\begin{eqnarray}
  F\out(y,\hat{z},\mu^z)
  &=& \frac{1}{p_{Y_i}(y)} \int_z z \, p_{Y_i|Z_i}(y|z) 
	\, \mc{CN}(z;\hat{z},\mu^z)  ,	\quad		\label{eq:Fout2}
\end{eqnarray}
where
$p_{Y_i}(y) \defn \int_z p_{Y_i|Z_i}(y|z) \mc{CN}(z;\hat{z},\mu^z)$.
From \eqref{pY|Z}, we rewrite $p_{Y_i|Z_i}(y|z)$ as
\begin{eqnarray}
  p_{Y_i|Z_i}(y|z)
  &=& \sum_{k=1}^{2^M} \frac{\beta_i\of{k}}{s\of{k}} 
	\,\mc{CN}\Big(z;\frac{y}{s\of{k}},\frac{\mu^v}{|s\of{k}|^2}\Big) ,
\end{eqnarray}
so that 
\begin{eqnarray}
  \int_z z\, p_{Y_i|Z_i}(y|z) \mc{CN}(z;\hat{z},\mu^z) 
  &=& \sum_{k=1}^{2^M} \frac{\beta_i\of{k}}{s\of{k}} \int_z z \,
	\mc{CN}\Big(z;\frac{y}{s\of{k}},\frac{\mu^v}{|s\of{k}|^2}\Big) 
	\mc{CN}(z;\hat{z},\mu^z) \quad \\
  p_{Y_i}(y) 
  &=& \sum_{k=1}^{2^M} \frac{\beta_i\of{k}}{s\of{k}} \int_z 
	\mc{CN}\Big(z;\frac{y}{s\of{k}},\frac{\mu^v}{|s\of{k}|^2}\Big) 
	\mc{CN}(z;\hat{z},\mu^z).
\end{eqnarray}
Using the property that
\begin{eqnarray}
  \mc{CN}(x;\hat{\theta},\mu^\theta)\mc{CN}(x;\hat{\phi},\mu^\phi)
  &=& \mc{CN}\Big(x;\frac{\hat{\theta}/\mu^\theta+\hat{\phi}/\mu^\phi}
   	{1/\mu^\theta+1/\mu^\phi},\frac{1}{1/\mu^\theta+1/\mu^\phi}\Big)
	\mc{CN}(0;\hat{\theta}-\hat{\phi},\mu^\theta+\mu^\phi) ,	
							\label{eq:pogr}
\end{eqnarray}
we can rewrite 
\begin{eqnarray}
  \lefteqn{ 
  \int_z z \, p_{Y_i|Z_i}(y|z) \, \mc{CN}(z;\hat{z},\mu^z) 
  }\nonumber\\
  &=& \sum_{k=1}^{2^M} \frac{\beta_i\of{k}}{s\of{k}} 
	\mc{CN}\Big(0;\frac{y_i}{s}-\hat{z},\frac{\mu^v}{|s\of{k}|^2}+\mu^z\Big)
	\int_z z \,
	\mc{CN}\bigg(z;\frac{
	  \frac{y}{s\of{k}}\frac{|s\of{k}|^2}{\mu^v}+\frac{\hat{z}}{\mu^z}
	}{
	  \frac{|s\of{k}|^2}{\mu^v}+\frac{1}{\mu^z}
	},\frac{1}{\frac{|s\of{k}|^2}{\mu^v}+\frac{1}{\mu^z}}\bigg)  
							\label{eq:prod}\\
  &=& \sum_{k=1}^{2^M} \frac{\beta_i\of{k}}{s\of{k}} 
	\mc{CN}\Big(\frac{y_i}{s};\hat{z},\frac{\mu^v}{|s\of{k}|^2}+\mu^z\Big)
	\frac{
	  \frac{y}{s\of{k}}\frac{|s\of{k}|^2}{\mu^v}+\frac{\hat{z}}{\mu^z}
	}{
	  \frac{|s\of{k}|^2}{\mu^v}+\frac{1}{\mu^z}
	} \quad \\
  &=& \sum_{k=1}^{2^M} \beta_i\of{k}
	\mc{CN}\big(y_i;s\of{k}\hat{z},|s\of{k}|^2\mu^z+\mu^v\big)
	\bigg( 
	\underbrace{ \Big(\frac{y}{s\of{k}}-\hat{z}\Big) 
		\frac{|s\of{k}|^2 \mu^z}{|s\of{k}|^2 \mu^z + \mu^v} 
	}_{\displaystyle \defn \hat{e}\of{k}(y,\hat{z},\mu^z)}
	+ \hat{z} \bigg)				\label{eq:intz}
\end{eqnarray}
and, using the same procedure, we get
\begin{eqnarray}
  p_{Y_i}(y) 
  &=& \sum_{k=1}^{2^M} \beta_i\of{k}
	\mc{CN}\big(y_i;s\of{k}\hat{z},|s\of{k}|^2\mu^z+\mu^v\big) .
							\label{eq:pY}
\end{eqnarray}
Finally, with $\xi_i\of{k}(y,\hat{z},\mu^z)$ defined in \eqref{xi}, 
equations \eqref{Fout2} and \eqref{intz} and \eqref{pY} combine to give
\begin{eqnarray}
  F\out(y,\hat{z},\mu^z) 
  &=& \sum_{k=1}^{2^M} \xi_i\of{k}(y,\hat{z},\mu^z)
	\big( \hat{e}\of{k}(y,\hat{z},\mu^z) + \hat{z} \big) ,	\quad
\end{eqnarray}
from which \eqref{Fout} follows immediately.

From (D1) and (D3), we have that
\begin{eqnarray}
  \mc{E}\out(y,\hat{z},\mu^z) 
  &=& \frac{ \int_z |z-F\out|^2 \, p_{Y_i|Z_i}(y|z) 
	\, \mc{CN}(z;\hat{z},\mu^z) }{ p_{Y_i}(y) } .	\label{eq:Eout2}
\end{eqnarray}
Similar to \eqref{prod}, we can write
\begin{eqnarray}
  \lefteqn{ \int_z |z-F\out|^2 \, p_{Y_i|Z_i}(y|z) 
	\, \mc{CN}(z;\hat{z},\mu^z) }\nonumber\\
  &=& \sum_{k=1}^{2^M} \frac{\beta_i\of{k}}{s\of{k}} 
	\mc{CN}\Big(0;\frac{y_i}{s}-\hat{z},\frac{\mu^v}{|s\of{k}|^2}+\mu^z\Big)
	\int_z |z-F\out|^2 \,
	\mc{CN}\bigg(z;\frac{
	  \frac{y}{s\of{k}}\frac{|s\of{k}|^2}{\mu^v}+\frac{\hat{z}}{\mu^z}
	}{
	  \frac{|s\of{k}|^2}{\mu^v}+\frac{1}{\mu^z}
	},\frac{1}{\frac{|s\of{k}|^2}{\mu^v}+\frac{1}{\mu^z}}\bigg) .
\end{eqnarray}
Then, using the change-of-variable $\tilde{z}\defn z-F\out$, and absorbing
the $s\of{k}$ terms as we did in \eqref{intz}, we get
\begin{eqnarray}
  \lefteqn{ \int_z |z-F\out|^2 \, p_{Y_i|Z_i}(y|z) 
	\, \mc{CN}(z;\hat{z},\mu^z) }\nonumber\\
  &=& \sum_{k=1}^{2^M} \beta_i\of{k}
	\mc{CN}\big(y_i;s\of{k}\hat{z},|s\of{k}|^2\mu^z+\mu^v\big)
	\int_{\tilde{z}} |\tilde{z}|^2 \,
	\mc{CN}\Big(\tilde{z}; \hat{e}\of{k}+
	\underbrace{\hat{z}-F\out}_{\displaystyle = -\hat{e}_i},
	\frac{\mu^v\mu^z}{|s\of{k}|^2\mu^z+\mu^v}\Big)  \\
  &=& \sum_{k=1}^{2^M} \beta_i\of{k}
	\mc{CN}\big(y_i;s\of{k}\hat{z},|s\of{k}|^2\mu^z+\mu^v\big)
	\Big( |\hat{e}\of{k}-\hat{e}_i|^2
	+ \frac{\mu^v\mu^z}{|s\of{k}|^2\mu^z+\mu^v} \Big).\label{eq:intz2}
\end{eqnarray}
Finally, using $\xi_i\of{k}(y,\hat{z},\mu^z)$ defined in \eqref{xi}, 
equations \eqref{pY} and \eqref{Eout2} and \eqref{intz2} combine to give
the expression for $\mc{E}\out(y,\hat{z},\mu^z)$ given in \eqref{Eout}.

\section{Derivation of RBP Quantities \textnormal{$F\inp$} and \textnormal{$\mc{E}\inp$}}
\label{app:in}

In this appendix, we derive the RBP quantities $F\inp(\hat{q},\mu^q)$ 
and $\mc{E}\inp(\hat{q},\mu^q)$ given in \eqref{Fin}-\eqref{nu}.

From (D4)-(D6), we note that $F\inp(\hat{q},\mu^q)$
and $\mc{E}\inp(\hat{q},\mu^q)$ are the mean and
variance, respectively, of the pdf 
\begin{eqnarray}
  \frac{1}{Z_j} p_{X_j}\!(q) \,\mc{CN}(q;\hat{q},\mu^q) ,\label{eq:pdfin}
\end{eqnarray}
where $Z_j=\int_q p_{X_j}\!(q) \,\mc{CN}(q;\hat{q},\mu^q)$.
Using \eqref{pogr} together with the definition of $p_{X_j}\!(.)$ 
from \eqref{pxj}, we find that
\begin{eqnarray}
  \lefteqn{
  p_{X_j}\!(q) \,\mc{CN}(q;\hat{q},\mu^q)
  }\nonumber\\
  &=& \lambda_j \mc{CN}(q;0,\mu_j) \,\mc{CN}(q;\hat{q},\mu^q) 
   	+ (1-\lambda_j) \delta(q) \,\mc{CN}(q;\hat{q},\mu^q) \\
  &=& 
	\lambda_j \mc{CN}(0;-\hat{q},\mu_j+\mu^q) \,
  	\mc{CN}\big(q;\frac{\hat{q}/\mu^q}{1/\mu_j+1/\mu^q},
		\frac{1}{1/\mu_j+1/\mu^q}\big) 
   	+ (1-\lambda_j) \mc{CN}(0;\hat{q},\mu^q) \delta(q)\\
  &=& 
  	\lambda_j \mc{CN}(\hat{q};0,\mu_j+\mu^q) \,
  	\mc{CN}\big(q;\frac{\hat{q}}{\mu^q}\frac{\mu^q\mu_j}{\mu^q+\mu_j},
		\frac{\mu^q\mu_j}{\mu^q+\mu_j}\big) 
   	+ (1-\lambda_j) \mc{CN}(\hat{q};0,\mu^q) \delta(q),
\end{eqnarray}
which implies that 
\begin{eqnarray}
  Z_j
  &=& \lambda_j \mc{CN}(\hat{q};0,\mu_j+\mu^q)
	+ (1-\lambda_j) \mc{CN}(\hat{q};0,\mu^q) .
\end{eqnarray}
Thus, the mean obeys
\begin{eqnarray}
  F\inp(\hat{q},\mu^q)
  &=& \frac{1}{Z_j}\int_q q \, p_{X_j}\!(q) \,\mc{CN}(q;\hat{q},\mu^q) \\
  &=& \underbrace{ \frac{\lambda_j}{Z_j}\mc{CN}(\hat{q};0,\mu_j+\mu^q) 
  	}_{\displaystyle =1/\alpha_j(\hat{q},\mu^q)}
	\underbrace{ \frac{\hat{q}}{\mu^q} \frac{\mu^q\mu_j}{\mu^q+\mu_j} 
	}_{\displaystyle =\gamma_j(\hat{q},\mu^q)}
							\label{eq:Fin2}
\end{eqnarray}
Expression \eqref{Fin} then follows directly from \eqref{Fin2}.

Since, for the pdf in \eqref{pdfin}, $F\inp$ is the mean and 
$\mc{E}\inp$ is the variance, we can write
\begin{eqnarray}
  \mc{E}\inp(\hat{q},\mu^q)
  &=& \frac{1}{Z_j}\int_q |q-F\inp|^2 \, p_{X_j}\!(q) 
  	\,\mc{CN}(q;\hat{q},\mu^q)  \\
  &=& \frac{1}{Z_j}\int_q |q|^2 \, p_{X_j}\!(q) 
  	\,\mc{CN}(q;\hat{q},\mu^q)  
  	-|F\inp|^2 \\
  &=& \frac{1}{Z_j}\int_{q} |q|^2 \, 
  	\Big(
	\lambda_j \mc{CN}(\hat{q};0,\mu_j+\mu^q) \,
  	\mc{CN}\big(q;
		\frac{\hat{q}}{\mu^q}\frac{\mu^q\mu_j}{\mu^q+\mu_j},
		\frac{\mu^q\mu_j}{\mu^q+\mu_j}\big) 
  	\nonumber\\&&\mbox{}
   	+ (1-\lambda_j) \mc{CN}(\hat{q};0,\mu^q) \delta(q)
	\Big)
  	-|F\inp|^2 \\
  &=& \underbrace{ 
  	\frac{\lambda_j}{Z_j} \mc{CN}(\hat{q};0,\mu_j\!+\!\mu^q) 
	}_{\displaystyle =1/\alpha_j}
  	\Big( |
	\underbrace{ 
		\frac{\hat{q}}{\mu^q}\frac{\mu^q\mu_j}{\mu^q+\mu_j}
	}_{\displaystyle = \gamma_j} |^2
	+ \underbrace{ 
	\frac{\mu^q\mu_j}{\mu^q+\mu_j}
	}_{\displaystyle = \nu_j} \Big) 
  	-|F\inp|^2 				\\
  &=& \frac{1}{\alpha_j}\big(|\gamma_j|^2+\nu_j\big)-\frac{1}{\alpha_j^2}
  	|\gamma_j|^2				\label{eq:Ein2}
\end{eqnarray}
Expression \eqref{Ein} then follows by rearranging \eqref{Ein2}.

\bibliographystyle{ieeetr}
\bibliography{macros_abbrev,books,misc,comm,multicarrier,sparse,stc}
\def\baselinestretch{1.0}
\end{document}